\documentclass[twocolumn,tighten,times,twocolappendix]{aastex631}

\usepackage{xspace}

\usepackage{comment}
\usepackage{amsmath}
\usepackage{CJKutf8}

\begin{document}


\title{Thick Disks around White Dwarfs viewed `Edge-off': Effects on Transit Properties and Infrared Excess}
\shorttitle{Transit and Infrared Excess from Thick Disks}
\shortauthors{Bhattacharjee}

\correspondingauthor{Soumyadeep Bhattacharjee}
\email{sbhatta2@caltech.edu}

%
%
%
%

\author[0000-0003-2071-2956]{Soumyadeep Bhattacharjee}
\affiliation{Department of Astronomy, California Institute of Technology, 1200 East California Blvd, Pasadena, CA, 91125, USA}

\begin{abstract}

A significant fraction of white dwarfs (WDs) host dust/debris
disks formed from the tidal disruption of asteroids and planetesimals. Several studies indicate that the disks can attain significant vertical heights through collisional cascade. In this work I model the effects of geometrically thick disks on two primary observables: photometric transits by the disk when viewed at high inclinations and infrared dust emission. Specifically, I consider disks with a Gaussian vertical profile with scale heights comparable to or larger than the WD radius. I primarily focus on inclinations $\gtrsim$$87$~degrees (`edge-off'), which can produce significant transits with moderate disk thickness. Both the transit depth and color become strong functions of inclination, and I explore their dependence on the disk parameters. I show that such a setup can produce the recently discovered reddening in the transit of WD\,J1013$-$0427. Moving to infrared emission, I show that the contribution from the heated inner rim can be substantial even at high inclinations. It can potentially explain the infrared excess observed in two transiting debris systems, WD\,1145$+$017 and WD\,1232$+$563, consistently with the transits. The other two important radiation components are the optically thin dust emission from the disk's outer layers and the optically thick emission from the backwarmed disk interior. Extending my analysis to G29-38 shows that the former can adequately produce the silicate emission feature with optically thin dust mass of $>$$10^{17}$~grams. The inner dense layers, on the other hand, allow the disk to contain orders of magnitude larger net dust mass. Overall, I show that thick disk effects can be significant and should be taken into account. I motivate detailed studies to quantify the effects accurately.

\end{abstract}


\section{Introduction}\label{sec:intro}

The past couple decades have seen significant improvements in our understanding of accretion of planetary remains onto white dwarfs. It is theorized that the primary formation channel of debris disks around white dwarfs is through tidal disruption of asteroids or planets which have been kicked into highly eccentric orbits entering the white dwarf's Roche potential \citep{Debes02, Jura03, Brouers22}. Over time, the material loses angular momentum through Poynting-Robertson drag or other, more efficient mechanisms, (see \citealt{Bonsor17,Brouwers22}) to finally fall on to the white dwarf's surface, resulting in photospheric metal pollution. This is a very common pathway, with $\gtrsim40\%$ of the white dwarfs being metal polluted \citep{Zuckerman03,Zuckerman10,Koester14,OuldRouis24}. 

Observations suggest that the in-falling planetary debris and dust form disks. The classic signature is infrared excess in the spectral energy distribution (SED) of white dwarfs arising from the heated dust emission. The first such system discovered is G29-38 \citep{Zuckerman87}, which is also the mostly widely studied system in this context. Since then, infrared excess has been detected in about $2\%$ of white dwarfs \citep{Mullally07, Jura07,Farihi08,Xu20,Wang23}. All of these white dwarfs also show metal pollution. 

Recently, mainly with the advent of large photometric surveys of the sky, another important signature has emerged - transits from the orbiting dust/debris. This is seen when the debris disk is viewed at sufficiently high inclinations. This was first detected in WD\,1145$+$017 by \cite{Vanderburg15}, with a dominant transit period of 4.5~hours, placing the debris at $\approx$$1\,R_{\odot}$ away from the white dwarf. Subsequently, transits with a wide range of periods have been detected: WD\,J0139+5245, with a period around 100\,days \citep{Vanderbosch20}, WD\,J0328$-$1219, with two detected periods of 9.94\,hours and 11.2\,hours \citep{Vanderbosch21}, WD\,1054$-$226, with a dominant period around 25\,hours \citep{Farihi22}, and WD\,J1944$+$4557 with a period of $4.9$~hours (Guidry and Vanderbosch et al., in prep). A potential period at $14.8$~hours was also recently detected in WD\,1232$+$563 \citep{Hermes25}. These measured orbital periods, with the exception of WD\,J0139+5245, places the debris at few $R_{\odot}$ distance from the white dwarf. There are nine more published systems without a measured orbital period \citep{Guidry21,Bhattacharjee25}. 

I now discuss the physical models that have been used to explain the above observables. The canonical setup to model the infrared excess assume a geometrically thin and optically thick (henceforth, `flat') configuration \citep{Jura03}. The model has been successfully used to explain the infrared SED in many white dwarfs and infer disk temperature and radius (see for example \citealt{Debes11}). But it has also been known for some time that a purely flat-disk model cannot explain several observations. For example, a flat disk cannot account for optically thin dust which is responsible for the dust emission features observed in many systems \citep{Jura07,Farihi18,Farihi25}. A flat disk also cannot explain the variability in infrared flux in several white dwarfs \citep{Xu18irvar, Swan20}. A possible solution is to have a second optically thin component to the disk (the two-component model, \citealt{Jura09,Xu18irvar}).

With regards to the high inclination (transiting) systems, a flat disk predicts very low infrared flux owing to the decreased projected disk emission area. This is consistent with at least two systems, WD\,J0328$-$1219 and WD\,1054$-$226, which have been shown to not have excess radiation in the infrared (although the former is uncertain, and may have some excess, see \citealt{Vanderbosch21}). However, two other transiting systems, WD\,1145$+$017 and WD\,1232$+$563 have significant infrared excess that cannot be explained by a flat disk (unless the disk is misaligned, \citealt{Izuierdo18}).\footnote{It would, however, be interesting to see how the two component model performs in this regard.} The remaining transiting systems do not have reliable infrared measurements.

Similar disk models have also been used to explain transits in the high inclination systems. The shape and contiguous nature of the transits often suggest that the transits are caused by clumps of dust or very small debris, rather than larger bodies \citep{Vanderburg15,Farihi22}. This indicates that the transit-causing and infrared-emitting disks can be related. Almost all of the detected transits are gray, i.e. the depths in different wavelengths are same. This has led to two model proposals. First is a diffused optically thin cloud which predicts absence of small sub-micron sized grains \citep{Alonso16}. The second is a flat disk viewed perfectly edge on \citep{Izuierdo18}. This allows for presence of small particles. Only very recently, significant reddening was detected in a long duration transit in WD\,1013$-$0427 \citep{Bhattacharjee25}. This disfavors a flat disk model, but can still be explained with a diffused optically thin cloud. However, it is difficult to explain the formation of such a cloud. Also, a purely optically thin cloud cannot be massive enough, whilst maintaining optical thin-ness, to account for the large amount of material accreted onto white dwarfs.

The above discussion can be summarized into one open question -- is it possible to form a disk model that can explain all the observations self-consistently? One possible solution is geometrically thick disks, which I explore in this work. Such disks naturally have both a diffuse optically thin and a dense optically thick component, which can potentially bridge the various observations. Note that it is conceptually similar to the two-component model, but the two components are now part of a single self-consistent entity.\footnote{In a way, the \cite{Jura09} model is equivalent to this setup when they present the optically thin component as a vertical extension to the flat disk.} Further motivation to consider such a disk comes from several recent works indicating that disks are expected to be geometrically thick due to collisional cascade \citep{Reach09, Kenyon17, Ballering22}, with scale heights comparable to or larger than the radius of the white dwarf.

This work attempts to model some key effects of a geometrically thick disk on transit and infrared observations. I primarily work in the high-inclination regime to try and relate transits properties and associated infrared excess. I mainly focus on three systems, WD\,J1013$-$0427 and the two transiting systems with infrared excess (WD\,1145$+$017 and WD\,1232$+$563), though I discuss generalizations to other systems whenever possible (including the applicability of my model to G29-38). The rest of the paper is as follows. In Section \ref{sec:transit_model}, I setup the transit calculation and investigate the transit properties, particularly the dependence of the depth and color on viewing angle. In Section \ref{sec:discussion}, I estimate the mid-infrared flux from radiation components which are natural consequences of thick disks but lacking in the flat disk model and compare with observations. Finally in Section \ref{sec:conclusions}, I present the limitations of my analyses and the conclusions.

\section{Transits with Thick Disks}\label{sec:transit_model}

The model below is a generalization of the transit models presented in previous papers \citep{Alonso16,Izuierdo18,Bhattacharjee25}. The key improvements that I aim for are: 1) lift any restriction on the optical depth of the disk, 2) include any particle size distribution, and 3) proper consideration of the detector bandpass. 

\begin{figure}[t]
    \centering
    \includegraphics[width=1.0\linewidth]{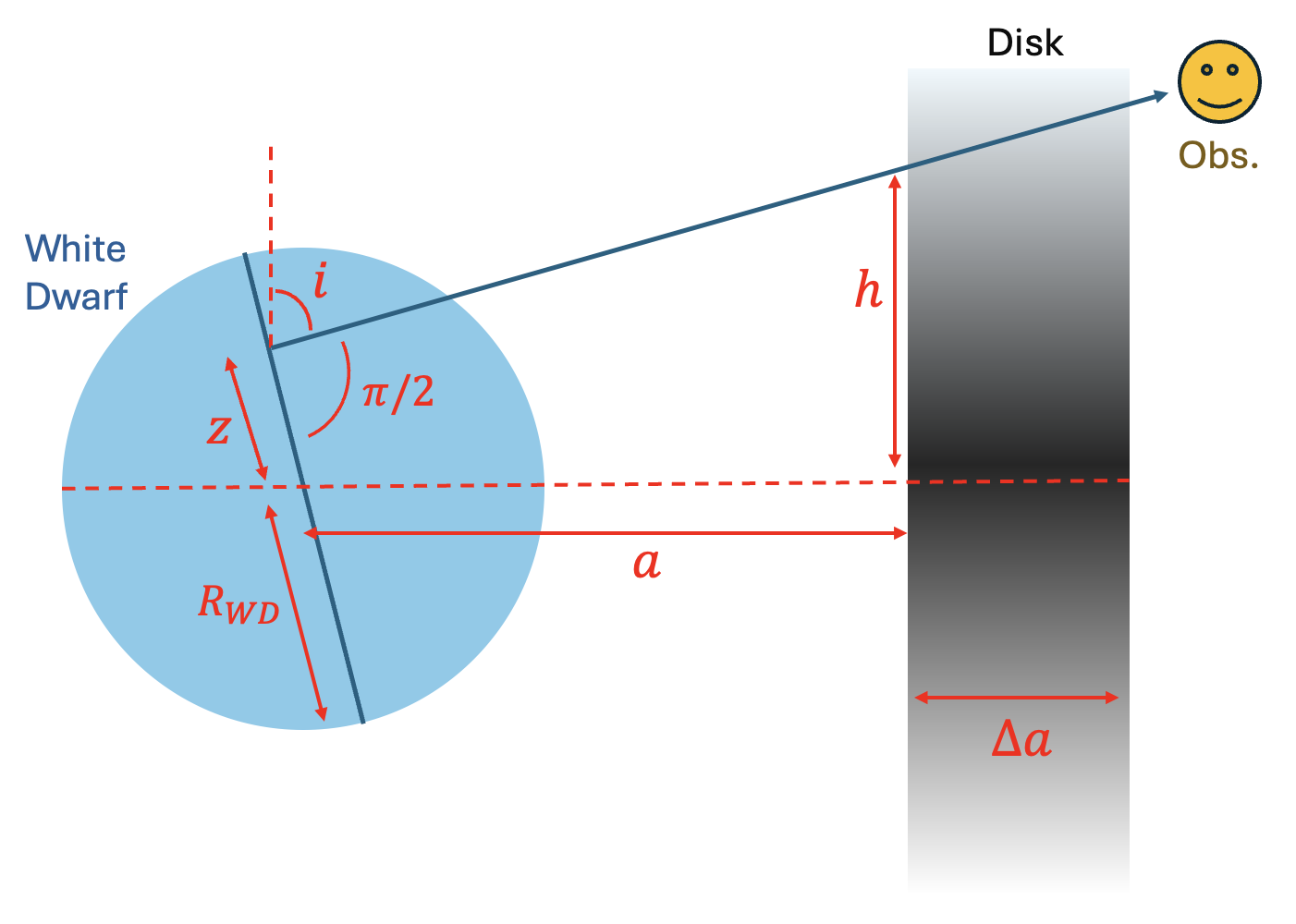}
    \caption{Cartoon (not to scale) showing the thick disk configuration and the observer line of sight responsible for the transits. All the disk parameters are marked. For the edge-on case, $z\equiv h$ which is discussed in Section~\ref{subsec:edgeon}. In the edge-off case, $f_1\equiv f_1(a,\,\Delta a,\,i,\,z)$ and is discussed in Section~\ref{sec:edgeoff}.}
    \label{fig:disk_schematic}
\end{figure}

A schematic of the setup which forms the basis of the following calculations is presented in Figure \ref{fig:disk_schematic}. I start by assuming the white dwarf as a circular projection on the sky. I define $z$ to be distance from the horizontal diameter of this circle as seen by an observer. The differential fractional area at a distance $z$ is then given by:
\begin{equation}
    \frac{dA(z)}{A_{\rm total}} = \frac{2dz\sqrt{R_{\rm WD}^2-z^2}}{\pi R_{\rm WD}^2} = \frac{2}{\pi}\sqrt{1-\left(\frac{z}{R_{\rm WD}}\right)^2}\frac{dz}{R_{\rm WD}}
\end{equation}
I assume that the white dwarf surface is uniformly lit with intensity $I_{\lambda,0}$ (thus ignoring limb darkening). The `flux' from the differential area will then be $I_{\lambda,0}dA(z)$. 

We now introduce the occulting dust disk. Assume a vertical disk profile of the form 
\begin{equation}
    N(h) = N_0f(h)
\end{equation}
where $N_0$ is the dust column density at $h=0$, where $h$ denotes the height from the disk plane, computed radially outward from the white dwarf. With this definition, any function with $f(0)=1$ is a valid disk profile. Following \cite{Izuierdo18}, I use a Gaussian profile for the rest of this work, given by \begin{equation}\label{eq:gaussian_profile}
    f(h) = e^{-\left(h/{h_z}\right)^2}
\end{equation}
where $h_z$ is the vertical scale height parameter. Note that by this definition, $h_z=\sqrt{2}\times$ the true scale height.

To model the extinction along this line of sight, I define the mean extinction cross section averaged over the particle distribution $n(r)$ (similar to as done in \citealt{Bhattacharjee25}):
\begin{equation}\label{eq:sigma_mean}
    \overline{\sigma}(\lambda) =  \frac{\int_{r_{\mathrm min}}^{r_{\mathrm max}}\sigma_{\mathrm ext}(r,\lambda)n(r)dr}{\int_{r_{\mathrm min}}^{r_{\mathrm max}}n(r)dr}.
\end{equation}
Now define $f_1(z)$ to be a function such that the optical depth traversed by the rays emitted from $z$ is $N_0f_1(z)\overline{\sigma}(\lambda)$. The relation between $f_1$, $h$, and $z$ depends on the inclination, $i$, and will be established in subsequent sections. The corresponding extincted `flux' from the differential area element is then: 
\begin{equation}
    dI(z)_{\lambda,\rm obs} = I_{\lambda,0}dA(z)e^{-N_0f_1(z)\overline{\sigma}(\lambda)}.
\end{equation}
I now define $\overline{\sigma}(\lambda) = \overline{\sigma}(\lambda_{\rm ref})g\left(\lambda/\lambda_{\rm ref}\right)$. This makes, $N_0\overline{\sigma}(\lambda_{\rm ref}) = \tau_{0,\rm ref}$, which is the peak optical depth at the reference wavelength. To account for the bandpass, I consider the filter transmission function of $\mathcal{T}(\lambda)$. I now define the depth in a bandpass as\footnote{The factor of $\lambda$ in the equation is to convert the intensity values to photon counts, which is what the detector measures. Note that an effective wavelength term (to convert the photon count to energy flux) gets eliminated while taking the ratios and, thus not included in the equations.}:
\begin{equation}
    1-D_{\rm band} = \frac{\int_{\lambda}\int_{z} dI(z)_{\lambda,\rm obs}\mathcal{T}(\lambda)\lambda d\lambda}{A_{\rm total}\int_{\lambda}I_{\lambda,0}\mathcal{T}(\lambda)\lambda d\lambda}
\end{equation}
This gives the observed depth as:
\begin{equation}
    1-D_{\rm band} = \frac{2}{\pi}\frac{\int_{-1}^{+1}\int_{\lambda} I_{\lambda,0}\sqrt{1-z'^2}e^{-\tau_{0,\rm ref}E(z',\lambda)}\mathcal{T}(\lambda)\lambda dz' d\lambda}{\int_{\lambda}I_{\lambda,0}\mathcal{T}(\lambda)\lambda d\lambda}
\end{equation}
where $z'=z/R_{\rm WD}$ and
\begin{equation}\label{eq:ez}
    E(z,\lambda) = f_1(z)g(\lambda/\lambda_{\rm ref})
\end{equation}
In the above two equations, $z$ is in the units of $R_{\rm WD}$ and $g(1)=1$. 

The wavelength dependence of the extinction, $g(\lambda)$ (the reference wavelength is dropped for ease of writing) is often parameterized as a power law:
\begin{equation}
    g(\lambda/\lambda_{\rm ref}) = (\lambda/\lambda_{\rm ref})^{-\alpha}
\end{equation}
where $\alpha$ is called the Angstrom exponent. I note that $g(\lambda)$ need not be a power law and can take any arbitrary form, which can be calculated using Equation \ref{eq:sigma_mean}. But a power law form is a good first-order approximation.

The exponent $\alpha$ can take a range of values, and it depends heavily on the particle size distribution. I briefly study this with two particle size distributions and varying the relevant parameter in Appendix \ref{app:angstrom_exp}. But overall it is quite improbable to achieve $\alpha\gtrsim3$ with any reasonable grain property and distribution. Thus, for my purpose, I work in the regime of $\alpha\leq3$.

The transit properties I aim to model in this work are the transit depth and color. The bandpass of the detector is an important aspect, as broader or overlapping bandpasses can dilute the transit colors significantly. In the recent past, the light curves from the Zwicky Transient Facility (ZTF) has led to the majority of the transit discoveries, including the first colored transit. Thus, for my purpose, I consider the depths in ZTF-$g$ and ZTF-$r$ bands (unless otherwise mentioned), which also closely resembles the photometric bands of the Sloan Digital Sky Survey. I thus define transit-color as the difference in the depth between the two bands: $D_g-D_r$. Without loss of generality, I take $\lambda_{\rm ref}=0.6\rm \mu m$ for the rest of this work. Also for simplicity (and preventing from being too specific) I assume the stellar spectra to be blackbody and transmission function to be a perfect top-hat (i.e. $\mathcal{T}=1$ inside the bandpass and $0$ otherwise). I assume approximate bandpasses of $0.42~\mu m-0.55~\mu m$ and $0.55~\mu m-0.72~\mu m$ for $g$ and $r$ bands, respectively.

\begin{figure*}[t]
    \centering
    \includegraphics[width=1.0\linewidth]{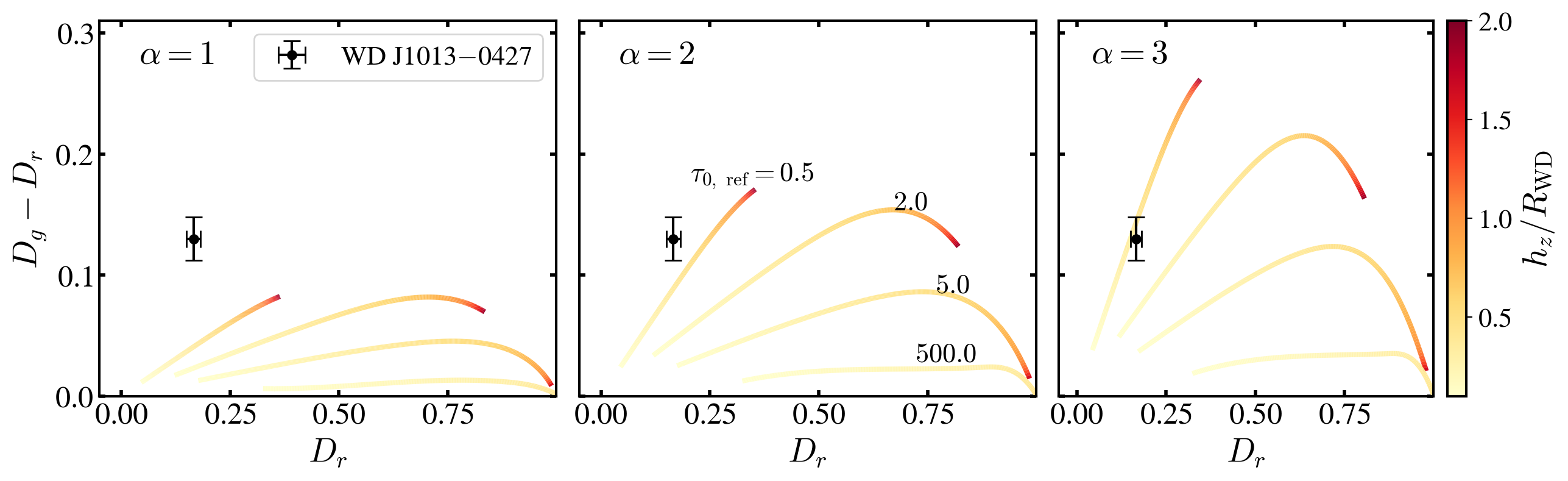}
    \caption{The color, $D_g-D_r$, as a function of the $r$-band depth, $D_r$ for a broad range of parameters for edge-on view. From left to right, the three values of $\alpha$ considered are mentioned in the figures. For each panel, I have considered four cases of $\tau_{0,~\rm ref}$, as mentioned in the second panel. For each optical depth, I consider a range of heights as indicated by the color-bar. The two primary takeaways are: 1) increasing $\alpha$ leads to increase in the color for a given depth, and 2) For a given optical depth and $\alpha$, there is a limit to the maximum color attainable. The depth and color of WD\,1013$-$0423 \citep{Bhattacharjee25} is marked in all the panels for comparison.}
    \label{fig:color_depth_space}
\end{figure*}

Unfortunately, there is not much clarity about the range of values for $h_z$ or $\tau_{0,~\rm ref}$. Also the past works which attempt to constrain these parameters do so for the dust disk from infrared excess, which may not be the same as the transit-causing disk. For my purpose, I adopt some fiducial range based on past works on both transits and dust disks (assuming they are not completely unrelated and share disk properties). For $h_z$, \cite{Izuierdo18} obtains a small $h_z/R_{\rm WD}\approx0.2$ for WD\,1145$+$017 (after proper translation of their parameters to ours) by fitting the transits with a flat disk model. On the other hand, \cite{Reach09} suggests a dust disk opening angle of $\gtrsim$$0.8$ degrees for G29-38 which, at a distance of $\sim$$1~R_{\odot}$ amounts to $h_z\gtrsim2~R_{\rm WD}$. Further radiative transfer modeling by \cite{Ballering22} also arrives at a similar limit. However, their best fit opening angle is even larger at $5$ degrees which yields $h_z\gtrsim10~R_{\rm WD}$. Based on these results, I work mostly in the conservative regime of $h_z/R_{\rm WD}\in[0.1,~2]$, unless otherwise mentioned.

Here, I briefly justify through simple kinematic argument why a scale height comparable to the white dwarf radius can be achieved through collisions. Assuming a vertical equilibrium between collisional dispersion and gravity, the disk scale height satisfies the relation $h_z = \sqrt{2}\sigma_v/\Omega_K$, where $\sigma_v$ and $\Omega_K$ are the velocity dispersion and Keplerian frequency, respectively. For $h_z=R_{\rm WD}$ at a distance of $a\approx1~R_{\odot}$ yields $\sigma_v\approx2.2~{\rm km~s^{-1}}$. Such a velocity dispersion can be achieved through small differential eccentricity, $e$, between the orbiting dust\slash debris. In small $e$ limit, the orbital speed $v_{\rm orb}\propto (1+e)$. At the assumed distance, $v_{\rm orb}\approx400~{\rm km~s^{-1}}$. Thus, a small $e$ differential of the order of $10^{-2}$ can yield the required velocity dispersion. Also, for given $h_z$, $\sigma_v\sim a^{-3/2}$ but $v_{\rm orb}\sim a^{-1/2}$. Thus it should be easier to attain high $h_z$ at larger distances. See also \cite{Swan21} for a similar argument.

The range of $\tau_{0,~\rm ref}$ is even less constrained than $h_z$, and it suffers transit with the disk structure given a total mass of the disk. \cite{Izuierdo18} obtains a wide range of lower limits from $\sim10$ to $\sim10^4$ for WD\,1145$+$017 for $\alpha\in[1,4]$\footnote{Note that \cite{Izuierdo18} Appendix A has an error of a factor of 2 in equation A1 (also noted in \citealt{Bhattacharjee25}). This propagates to A9, where term in the exponential should be twice (note another typo here, where the ratio inside the square root should be squared). This increases their estimates of lower limit of $\tau$ by two orders of magnitudes. I report the optical depth limits after correcting for this errors.}. On the other hand, the model by \cite{Reach09} for G29-38 uses $\tau\lesssim10$. I thus consider a broad range of values from less than unity to few thousands.~\footnote{In my model, the total mass in the disk is approximately:
\begin{equation}\label{eq:mdust}
    M_{\rm dust} = m_{\rm grain}N_{\rm dust}\approx m_{\rm grain}\frac{\tau_{0,~\rm ref}}{\overline{\sigma}_{\rm ref}}2\pi a\int_{-\infty}^{\infty}f(z)dz
\end{equation}
where $m_{\rm grain}$ is the mean mass of the grain, $N_{\rm dust}$ is the total number of dust particles, $\overline{\sigma}_{\rm ref}$ is the mean cross section at the reference wavelength, and $a$ is the distance from the white dwarf. Assuming a fiducial grain radius of $1~\rm \mu m$, $a\sim1~R_{\odot}$, and silicate density of $2~\rm g~cm^{-3}$, a mass of $M_{\rm dust}\approx10^{20}~g$ yields $\tau_{0,~\rm ref}\approx10^3$.}

\subsection{Viewing Edge on}\label{subsec:edgeon}

I first consider the simple case of viewing the disk perfectly edge on (i.e. inclination $i=90$ degrees). Fixing the inclination enables better understanding of the effect of the disk parameters: $\alpha$, $h_z$, and $\tau_{0,~\rm ref}$, on the transit properties. In this case, the line of sight is parallel to the disk plane. Thus, $z= h$ and $f_1(z[h])=f(h)$. I consider WD\,J1013$-$0427 as my reference object for color in transit. Thus, I use its temperature of $21,900$~K to perform the calculations. Manual inspection shows that the results do not change much with other temperatures.

The resultant color as a function of depth is shown in Figure \ref{fig:color_depth_space}. There are two main takeaways from this figure. The first is the expected behavior that increasing $\alpha$ results in increasing color for a given transit depth. The second and more interesting feature is the variation of color with $\tau_{0,~\rm ref}$ and $h_z$. For a given $\tau_{0,~\rm ref}$ with increasing $h_z$ initially the color and depth both increases. However, a maximum color is then attained, after which the color decreases though the depth keeps on increasing (this is only seen for high enough $\tau_{0,~\rm ref}$ values, as for low values the depth is limited by the maximum attenuation possible). This is understood as follows. In the low $h_z$ regime, the color is produced by the effective vertical height of the disk for which it is optically thick to the radiation of the band in concern (the limiting case for which gives the \citealt{Izuierdo18} model). But if the scale height is very low, all the dust is concentrated in the mid-plane making the optical depth gradient too high for it to create any difference in the $\tau=1$ height for the different wavelengths. On the other hand, if $h_z$ is very high, the whole of the white dwarf is uniformly occulted by dust, thus decreasing the color. Thus, there is a sweet spot in between where there is a maximum color in the dust transit.\footnote{An easy mathematical viewpoint is to consider the behavior of the function $e^{-x}-e^{-y}$ when $y\geq x$, which denotes the transit color ($x$, $y$ being the band optical depths). The function $\rightarrow0$ both when $x,y\rightarrow0$ and $x,y\rightarrow\infty$. Thus, there ought to be a maxima.}

For comparison, I mark the depth and color of WD\,J1013$-$0427 in all the panels. It is clear that only a high $\alpha\approx3$ value and sufficiently low optical depth can simultaneously produce the ZTF depth and color. This is in agreement with that inferred from studying the two limiting models in \cite{Bhattacharjee25}. Also, such a high value of $\alpha$ can only be achieved with very small dust grains (see Appendix \ref{app:angstrom_exp} and \ref{fig:angstrom_exponent}). This is, thus, an indication of an optically thin (or, at least non-thick) component to the dust occulting the white dwarf. This, however, may not imply that the whole disk is optically thin. This is addressed in the next section.

\subsection{Viewing Edge-off}\label{sec:edgeoff}

It is easy to visualize that, for a thick disk, dust transit can occur even when the inclination is not completely edge-on. The transit is then caused by the outer layers of the disk. But, owing to the vertical gradient, these layers are much less dense than the mid-plane and thus will lead to different transit behavior. To study this effect, I now develop a more general formalism for slight `edge-off' lines of sight. In this scenario, I need the knowledge of two additional properties of the disk: the distance of the inner edge from the white dwarf, $a$, and the width of the disk, $\Delta a$. For a given inclination $i$, the line of sight traverses a vertical segment through the disk of length $\Delta a\cot{(i)}$. Thus, for such a segment with a height $h$ above the disk plane at the inner edge, the total optical depth along the line of sight is given by:
\begin{equation}\label{eq:inclined_tau}
    \tau_{\rm ref}[h(z)] = \frac{\tau_{0,~\rm ref}}{\Delta a}\int_{h(z)}^{h(z)+\Delta a\cot(i)}f(x)\frac{dx}{\cos(i)},
\end{equation}
where $h(z)$ denotes the height in the disk which gets projected onto the white dwarf at $z$, and $x$ is the dummy integral variable. A simple geometric calculation yields:
\begin{equation}\label{eq:projection}
    h(z)= a\cot(i)+z/\sin(i).
\end{equation}
Thus, I have $f_1(z) = \tau_{\rm ref}[h(z)]/\tau_{0,~\rm ref}$, which I use in Equation~\ref{eq:ez}.\footnote{Note that in the limit of $i\rightarrow90$~degrees, this formula correctly approaches $\tau_{\rm ref}(z) = \tau_{0,~\rm ref}f(z)$, which is the edge-on case. } This model is, in general, valid for any inclinations unless the inclination is too low that the plane-parallel vertical-only structure of the disk is no longer a valid~assumption. 

\begin{figure}[t]
    \centering
    \includegraphics[width=1.0\linewidth]{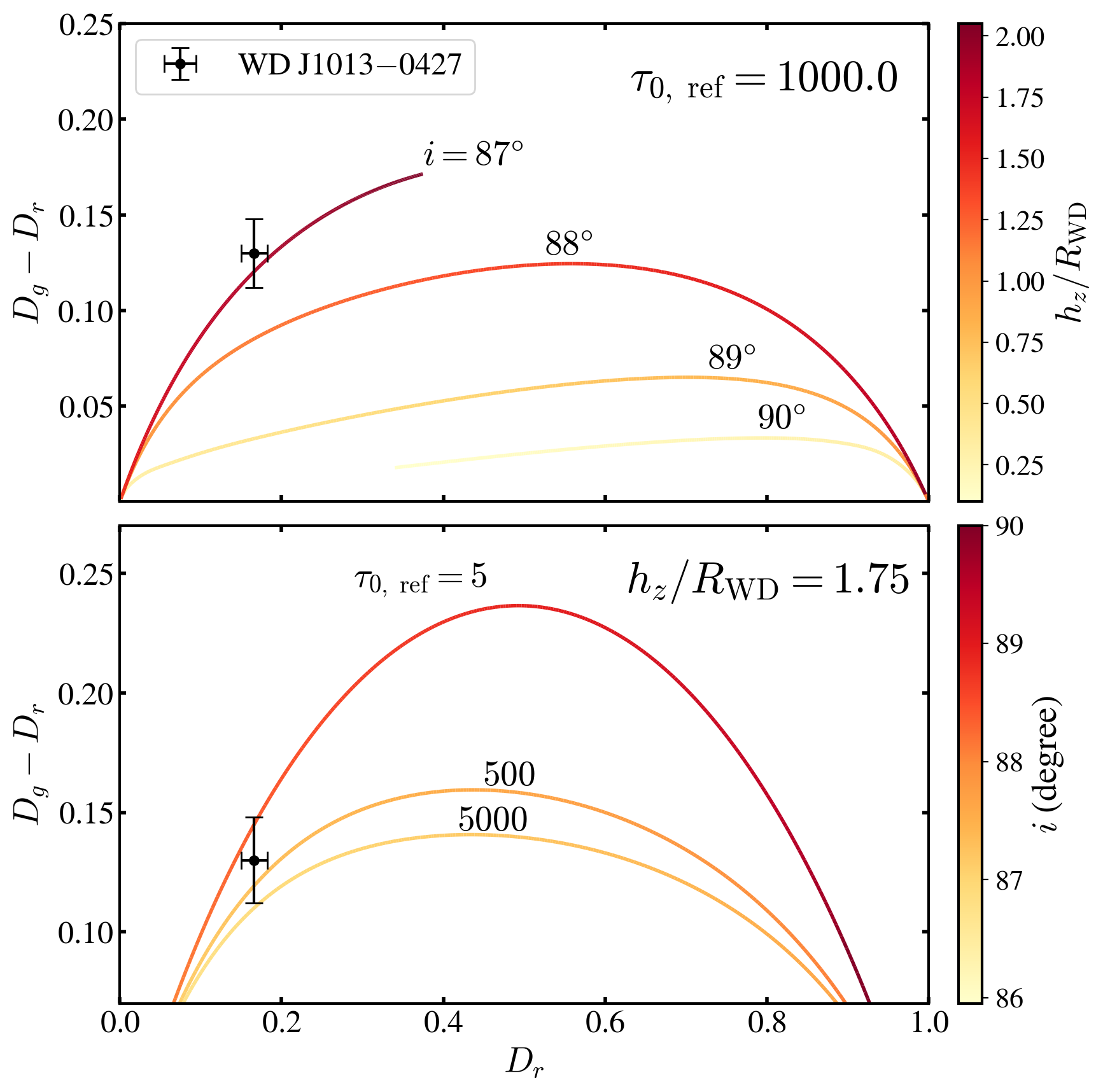}
    \caption{The color as a function of depth for fixed $\tau_{0,~\rm ref}$ (top panel) and $h_z$ (bottom panel) when $a=100~R_{\rm WD}$ and $\Delta a=10~R_{\rm WD}$ are assumed. The parameters being varied are specified as in-figure texts and the colorbar. In all the cases, $\alpha$ is fixed at $3$ to achieve the maximum color-dependence. It is evident that edge-off viewing can achieve significant color, given the scale height is sufficiently large. The position of WD\,J1013$-$0427 is marked in both panels for comparison. The transit properties of this object can indeed be explained with edge-off view of a geometrically thick disk with a high mid-plane optical depth. The results remain qualitatively the same for larger $a$, but the allowed inclination angles become limited (see Appendix~\ref{app:transit_a_dep}). }
    \label{fig:inclination_effect}
\end{figure}

\subsubsection{WD\,J1013$-$0427}\label{subsubsec:wd1013}

To demonstrate the effect of $i$, I fix few of the other parameters and, once again, refer to WD\,J1013$-$0427 for comparison. I take $\alpha=3$, to maximize the possibility of getting a color. I use a white dwarf temperature of $21,900$~K to resemble WD\,J1013-0427, but the effect is the same for any other temperature. The distance to the disk is unknown. For most of the objects with detected transit periods, the corresponding orbits lie in the approximate range of $100-300~R_{\rm WD}$. Here, I consider the case of $a=100\,R_{\rm WD}$ (for the ease of the readers, to be consistent with the parameters assumed for WD\,1145$+$017 in the following sections). But, as discussed later in Section~\ref{subsec:reconciling_with_transits}, this is unlikely to be the case and the disk is expected to be farther away. The only effect of placing the disk farther away is to narrow the range of possible edge-off viewing angles. This is shown in Appendix \ref{app:transit_a_dep}. The width of the disk is also unconstrained. We adopt a fiducial value of $\Delta a=10~R_{\rm WD}$ (motivated from the upper limit of the dust disk in \citealt{Ballering22}), but the effect of this parameter is discussed shortly.

First, I fix $\tau_{0,~\rm ref}$ to be constant at a fairly high fiducial value of $10^3$ and vary $i$ and $h_z$ (top panel of Figure \ref{fig:inclination_effect}). I see that it is possible to increase the amount of color in the transit by decreasing $i$. Intuitively this is because the outer layers of the disk are less dense, and thus mimics the effect of an optically thinner disk producing color in the transit. The `trade-off' is the requirement a high enough $h_z$ so that the transits are produced at all. 

I now compare the effect of different mid-plane optical depths. To do this, I fix $h_z=1.75$ to vary $i$ and $\tau_{0,~\rm ref}$ (bottom panel of the same figure). I see that the data for WD\,J1013-0427 can be fit satisfactorily with a broad range of $\tau_{0,~\rm ref}$ with edge-off viewing angles. 

\begin{figure}
    \centering
    \includegraphics[width=1\linewidth]{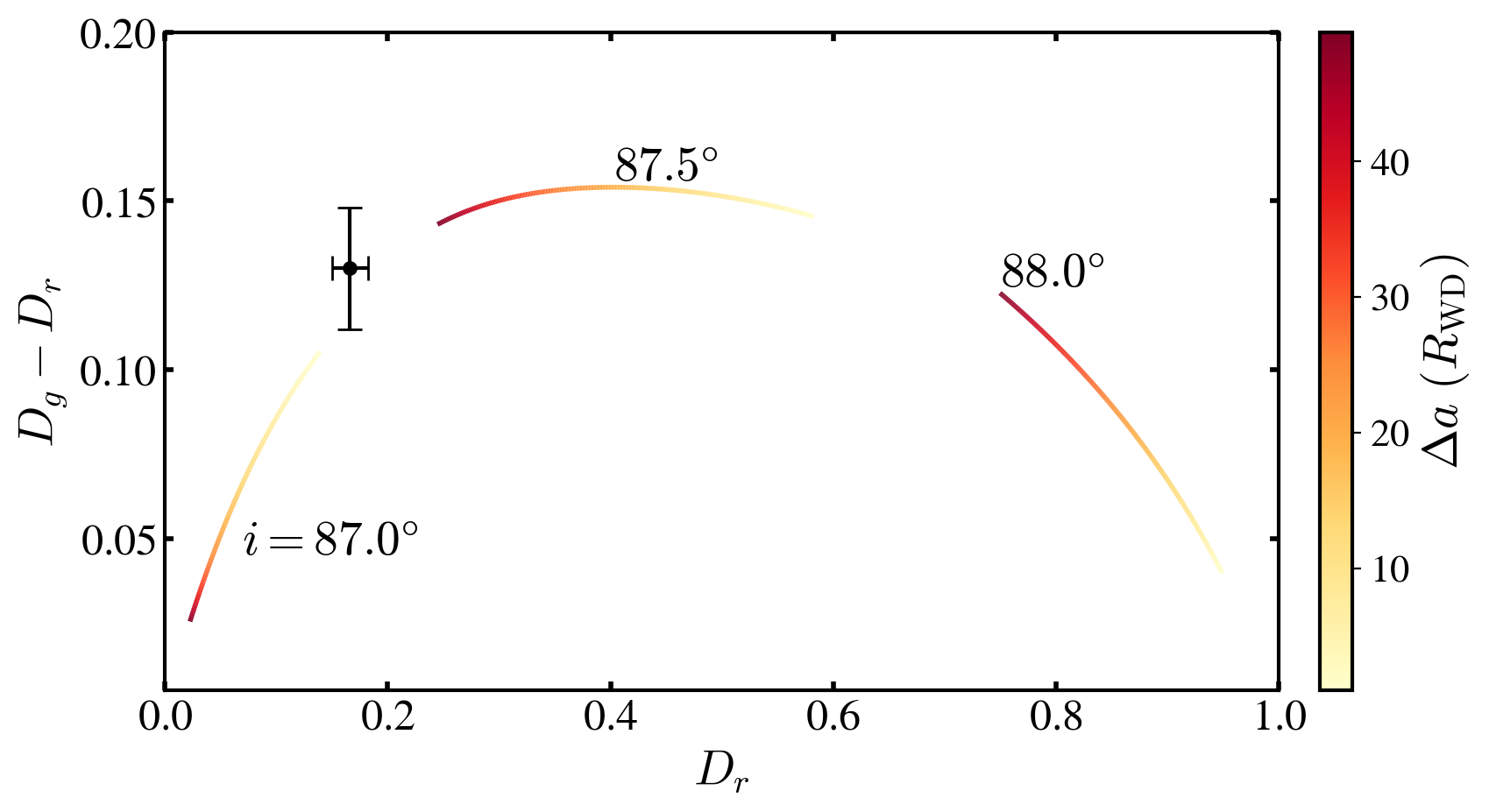}
    \caption{The effect of $\Delta a$ in the transit depth and color. In this figure, we fix the other parameters at $a=100~R_{\rm WD}$, $h_z/R_{\rm WD}=1.75$, and $\tau_{0,~\rm ref}=10^3$. $\Delta a$ is varied from $1~R_{\rm WD}$ to $50~R_{\rm WD}$. WD\,J1013$-$0427 is marked for comparison. Lower inclination requires a smaller $\Delta a$ to satisfy the object's transit properties.}
    \label{fig:deltaa_effect}
\end{figure}

The dependence on $\Delta a$ is more complicated. We demonstrate the effect of this parameter in Figure \ref{fig:deltaa_effect} by fixing $a=100~R_{\rm WD}$, $h_z/R_{\rm WD}=1.75$, and $\tau_{0,~\rm ref}=10^3$, but varying $\Delta a$ for three $i$. We see that for all $i$, the depth decreases with increasing $\Delta a$. This is because, as $\Delta a$ increases, the line of sight traverses more of the outer layer of low optical depth. The color variation, however, is dependent on $i$. For lower $i$, the color is limited by the maximum depth achievable which decreases with increasing $\Delta a$. For higher $i$, on the other hand, the depth is sufficient for the color to increase with increase in the amount of optically thin dust along the sightline. This introduces an additional degeneracy in the model (on top of that from $h_z$, as presented earlier) as the same depth and color can be achieved through either changing $\Delta a$ or the inclination angle. It is thus important to consistently analyze other observables (like infrared excess, see Section~\ref{sec:discussion}) to break the degeneracies.

An additional unique aspect of WD\,J1013$-$0427 is that it is the only transiting system known showing double-peaked gas emission from a disk. This is also an indication of edge-off viewing as, for egde-on views, one would expect to see gaseous absorption lines instead. \cite{Bhattacharjee25} uses appropriate emission model and arrives at a gas disk inclination of $\gtrsim80$~degrees. Thus, the observations are consistent with the coplanarity of the transit-causing dust\slash debris disk and the gas~disk.

\subsubsection{Gray Transitors}\label{subsubsec:gray_transitors}

The transiting systems other than WD\,J1013$-$0427 are gray, with only upper limits on the colors. For example, \cite{Izuierdo18}, using spectrophotometric observations performed with the Optical System for Imaging and low-Intermediate-Resolution Integrated Spectroscopy on the Gran Telescopio Canarias (GTC) infers the upper limit for WD\,1145$+$017 to be $D_{g,~\rm GTC}-D_{r,~\rm GTC}\lesssim0.06$.\footnote{Note that the GTC passbands they used are significantly different from that of ZTF, thus I use the photometric band passes as defined in \cite{Izuierdo18} in their section 3.3.} The transit depths, however, were significant and mostly in the range of $\approx$$20\%-50\%$. Recently, \cite{Hermes25} analyzed WD\,1232$+$563 and arrived at a limit of $\Delta(m_g-m_r)<0.024\implies D_g-D_r\lesssim0.02$, with similar transit depths as WD\,1145$+$017. As mentioned earlier, I specifically consider these two systems in my work because they also have detected infrared excess. But the other gray systems have similar color limits (see for example \citealt{Vanderbosch21} or Guidry and Vanderbosch et al. in prep). For the ease of discussion, I will focus on WD\,1145$+$017 in this section (and move the discussion on WD\,1232$+$563 to Appendix \ref{app:wd1232}; the results are similar). 

\begin{figure}[t]
    \centering
    \includegraphics[width=1.0\linewidth]{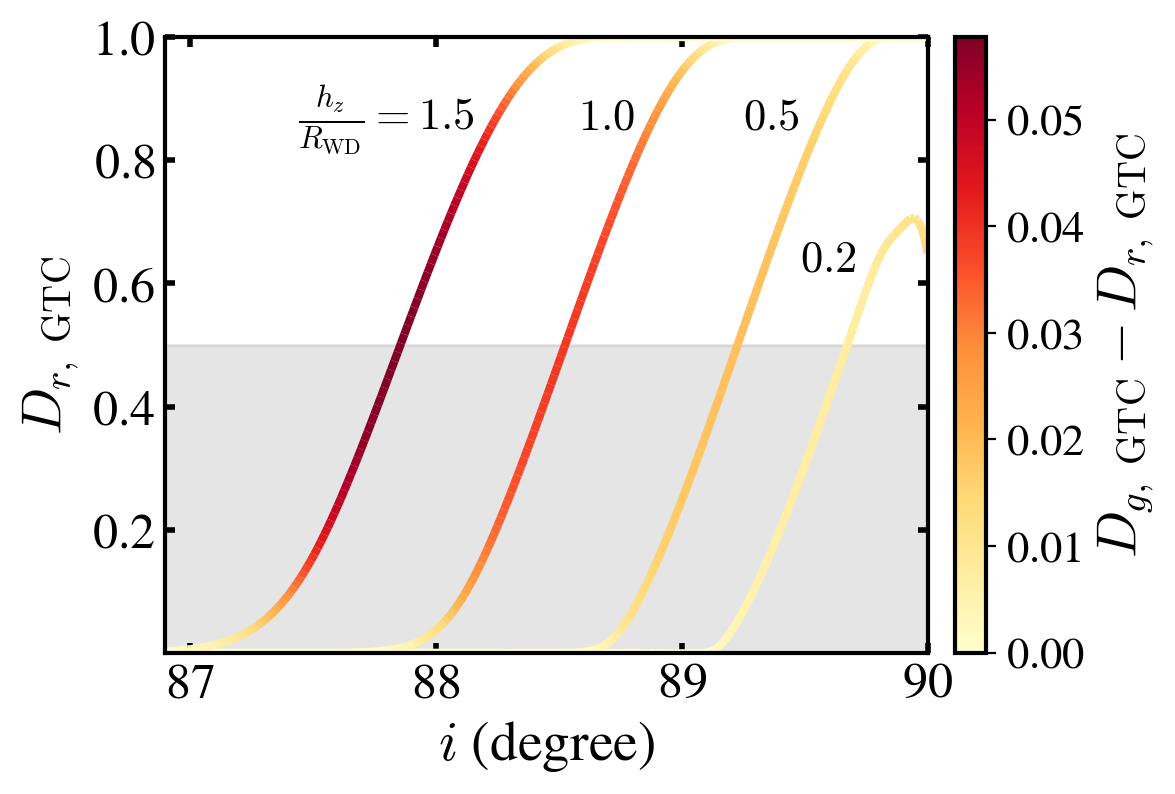}
    \caption{The depth as a function of inclination for $\alpha=0.5$, $\tau_{0,~\rm ref}=10^3$, and $\Delta a=10~R_{\rm WD}$ with the stellar parameters of WD\,1145$+$017. The four disk heights considered are labeled in the figure and the corresponding curves are colored according to the transit color. The shaded region marks the range of transit depths observed. The transit colors are all within the upper limit derived in \cite{Izuierdo18}, still maintaining the observed depths. Lowering $\alpha$ (or increasing $\tau_{0,~\rm ref}$, though this effect is marginal) enables broader range of edge-off inclination. Increasing $\Delta a$ effectively shifts the curves rightward while slightly increasing the maximum~color.}
    \label{fig:inclination_effect_colorless}
\end{figure}

First, I briefly review the past attempts to model the transits in WD\,1145$+$017. \cite{Alonso16} used an optically thin model and arrived at the constraint of $\alpha\approx0$ (implying a lack of small grains). Such a model is unlikely as it is difficult to explain the formation of a purely diffused and optically thin disk. On the other hand, \cite{Izuierdo18} invokes a perfectly edge-on optically thick and geometrically thin disk. This explains the transit properties well for any assumed $\alpha$, but fails to explain the infrared excess without invoking a misaligned dust disk. Additionally, a perfectly edge-on inclination is statistically improbable.

Interestingly, both the above models are two extreme cases of my model. However, owing to their individual limitations, I explore the remainder of the parameter space, especially focusing on the possibility of edge-off viewing angles. The transit period of the object\footnote{At the time the transits were active, that is. Currently, the object does not show photometric transits.} was $4.5$ hours which translates to $a\approx100~R_{\rm WD}$, already in line with my previous fiducial case. The extent of the disk is not well constrained, so I continue to use my fiducial case of $\Delta a=10~R_{\rm WD}$. Inspection shows that increasing $\Delta a$ (even up to $\approx$$100~R_{\rm WD}$) does not change our subsequent discussion, thus we stick to our fiducial assumption for the rest of the section. I use $15020$~K as the white dwarf temperature from \cite{Izuierdo18}. 

Manual inspection shows that with $\alpha\in[0,~1]$, it is possible to remain gray (i.e. well within the color limit set by \citealt{Izuierdo18}) for edge-off inclination angles. I demonstrate this in Figure \ref{fig:inclination_effect_colorless} by assuming $\alpha=0.5$ (approximately the limiting value for power law distribution of grain size, see Appendix \ref{app:angstrom_exp}) and $\tau_{0,~\rm ref}=10^3$. This clearly shows that the observed transit properties can be satisfied under a much broader range of parameters, thus significantly relaxing the stringent conditions imposed in the previous models. Specifically, it opens up the possibility of edge-off viewing angles, which, as I shall show in the next section, may have significant implications on the understanding of the infrared excess. A further dearth of small particles will result in lower $\alpha$ which nothing but aids my discussion. Higher values of $\alpha$ (i.e. allowing for an abundance of small particles), however, requires almost perfectly edge-on view and very thin disks to remain within the color limits (thus approaching the \citealt{Izuierdo18} model). Given that such a configuration is statistically unlikely, my model favors a low $\alpha$ and, thus, in agreement with the conclusions of \cite{Alonso16,Xu18}.

\subsubsection{Colored vs Gray: Is WD\,J1013$-$0427 special?}

Unifying WD\,J1013$-$0427 with the gray transitor majority is not straightforward. As shown, a high $\alpha\approx3$ is required to explain the color of WD\,J1013$-$0427. For the gray transitors, similar $\alpha$ value would mean almost perfectly edge-on view. One can then conjecture that WD\,J1013$-$0427 is a system viewed edge-off, whereas the gray transitors are edge-on systems. However, given that only one out of fourteen transiting system (till date) shows color, such a scenario is not consistent with a random distribution of viewing angles. 

The other scenario might be that all the objects are indeed edge-on systems with disks of small scale heights. Even in this regime, WD\,J1013$-$0427 appears an outlier as one would require a low value of $\tau_{0,~\rm ref}\approx1$ (see Figure \ref{fig:color_depth_space}) to explain its color, unlike the gray systems. 

The remaining case of all systems being edge-off is statistically most likely. But even this would point to the uniqueness of WD\,J1013$-$0427 with having an abundance of small particles (required for the high $\alpha$ value) unlike the gray transitors. 

One can, however, reconcile the difference with the following scenario. Most of the previous detections of transiting systems have been biased towards objects showing consistent transits. I think that these are systems with relatively higher $i$ and the search has been biased to these systems. This would mean that there are more systems where our line of sight is edge-off, but owing to lack of continuous transits they are not often detected.\footnote{Indeed several long-duration transit events are being detected in long baseline surveys like ZTF, see  \citealt{Bhattacharjee25} and van Roestel et al. in prep.} WD\,J1013$-$0427 can be one of them, but possibly with a collision event which significantly enhanced the disk scale height (i.e. increasing $h_z$), and production of small particles. However, the associated timescales remain to be explained consistently. For example, for WD\,J1013$-$0427, a shorter egress than ingress is against the general picture of collisions. 

\subsection{Achieving the transits in light curves}

We now discuss an important caveat of the model, which assumes that the disk is uniform and static. But this predicts a single value of the attenuated flux of the white dwarf, and cannot produce the transits in the light curve. No uniform disk model can produce the time-dependent transits. To achieve that, the disk is required to be non-uniform and also evolve over time. But this is hardly surprising as the disk is a very dynamic environment with processes like collisions, dispersion and dissipation \citep{Swan20}, tidal disruption \citep{Veras17} and volcanism \citep{Li25} ongoing continuously. This will inevitably create structures within the disk -- dust clumps, varying disk thickness, etc. Incorporating these effects would require dynamical simulations (similar to \citealt{Veras17}) along with radiative transfer calculations, which is beyond the scope of this work.

Nevertheless, I now argue that my model is parametrically consistent with the above picture. I firstly note that the disk circumference in my model is much (at least a factor of $\sim$$100\pi$) larger than the white dwarf dimensions. Thus, the analyses in the previous sections can be regarded as those of `snapshots' of the disk as it orbits the white dwarf. The non-uniformnesses can then be captured by considering orbital phase-dependent values for disk parameters. One crucial parameter is $h_z$. Figure~\ref{fig:inclination_effect_colorless} shows that the transit depth is a strong function of $h_z$. Thus, relatively small variation of the disk's vertical extent over its circumference (potentially through different levels of collisional activities) can produce the transits over one orbital period. For example, inspection shows that a moderate variation of $h_z$ in the range of $0.7-1.2~R_{\rm WD}$ over the disk circumference can explain the short-term transits in WD\,1145$+$017. Additional contribution can also arise from varying $\tau_{0,~\rm ref}$, which may parametrically represent over and under-dense regions of the disk. \footnote{Varying disk radial extent can also, in principle, contribute to the transits. However, how a significant variation in $\Delta a$ across the circumference can be maintained is unclear. Finally, changing inclination can also result in long-duration variation in transit activity through disk precession.}

Note that this also naturally explains the variability in transit activity observed, both on short and long timescales. When the disk is collisionally perturbed, it achieves a higher scale height of dust/debris yielding larger depths. Over time the scale height decreases as the disk settles, leading to decrease in the transit activity. I note that this also predicts that the period of transits remain nearly equal each time the transits resume, as the orbit of the disk is not changed. This is at par with what has been observed recently in a transiting debris system (WD\,J1944, Guidry and Vanderbosch et al. in prep). Recently, WD\,1145$+$017 has stopped transit activity. The period of the resumed activity at any later time will further test this theory.

\section{Infrared Emission from Thick Disks}\label{sec:discussion}

A joint study of the transits and infrared excess would provide novel constraints on disk structure. Thus, in this section I investigate the infrared radiation signatures from the thick disk model formulated in the previous sections. I note here that there is no convincing reason to believe that the infrared emitting dust disk and the transit causing debris disk are the same. Nevertheless, for the purpose of this work, I consider a similar range of disk parameters as in Section \ref{sec:transit_model}. As I shall then show that, within my model, it is possible to explain both the transits and the infrared excess in edge-off systems with a single consistent disk.

I start with a brief review of the classical flat disk model in Section \ref{sec:flat_disk}. In Section~\ref{sec:ir_thick_disk}, I discuss the thick disk radiation, and compare with the classical model whenever applicable. The thick disk calculations are motivated largely by several past works on protoplanetary disks (including, but not limited to, \citealt{Calvet91,Chiang97,Dullemond03,Kama09,Dullemond10,Flock16}). For the purpose of this work, I use simple analytic prescriptions to demonstrate the importance of several effects lacking in the classical flat disk model. I expect that this will motivate more detailed analyses to gauge the importance of each of these effects individually.

\subsection{Reviewing the Classical Flat-Disk Model}\label{sec:flat_disk}

The most widely used model to describe infrared flux from dust disks around white dwarfs was formulated by \cite{Jura03}. The model is derived from the discussion in \cite{Chiang97} and assumes a flat and optically thick disk model. The flux density, $F_{\nu}$, at inclination $i$ and distance $d$ is given by:
\begin{equation}
    F_{\nu} = \frac{2\pi\cos(i)}{d^2}\int_a B_{\nu}(T_i)a\,da
\end{equation}
where $a$ is the distance to the disk element from the white dwarf, and $T_i$ is the disk internal temperature is given by:
\begin{equation}\label{eq:temp_jura}
    T_{i} = \left(\frac{2}{3\pi}\right)^{1/2}\left(\frac{R_{\rm WD}}{a}\right)^{3/4}T_{\rm WD},
\end{equation}
and the integration is carried over the disk radial extent. The model has been widely successful in explaining the infrared signatures from many white dwarf systems. However, as mentioned in Section~\ref{sec:intro}, it is difficult to apply this model to high inclination systems. The model requires the disk to be placed very close to the white dwarf to achieve the color temperature of the infrared SEDs. This reduces the emission surface area (the outer radius can be much larger, but the flux is subdominant due to low temperatures given by Equation~\ref{eq:temp_jura}). The projected emission area at high inclinations becomes very small, yielding very low flux. Thus the flat disk cannot produce the observed infrared flux in the two transiting systems, WD\,1145$+$017 and WD\,J1228$+$1040, unless the it is significantly misaligned with the transit-causing debris disk, \citep{Izuierdo18}. In the next few sections, I attempt to resolve this issue with my thick disk model.

\subsection{Irradiated Thick Disk}\label{sec:ir_thick_disk}

\begin{figure}[t]
    \centering
    \includegraphics[width=1.0\linewidth]{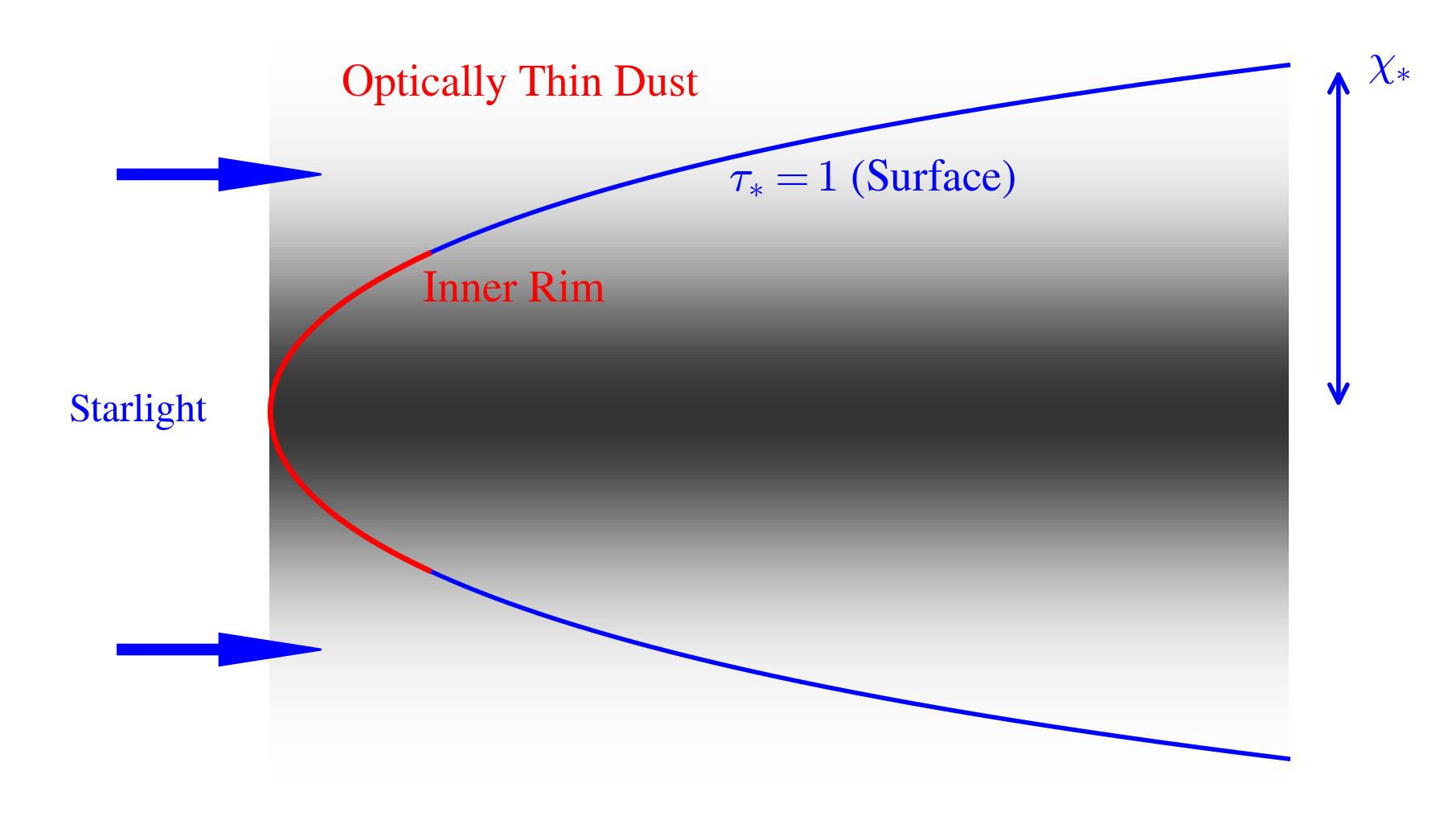}
    \caption{Cartoon figure showing the thick disk being irradiated by the white dwarf. The gray scale shows the vertical Gaussian density profile. The surface of the disk is defined by the locus of points where the optical depth of a radial stellar photon is unity ($\tau_{*}=1$). The dust outside the surface is optically thin and directly heated by the white dwarf to temperatures $\gtrsim10^3$~K, and thus responsible for the optically thin emission. A section of the surface, the inner rim, behaves like a optically thick wall emitting at similarly high temperatures. This potentially contributes significantly to the near to mid infrared continuum.}
    \label{fig:irradiated_disk}
\end{figure}

I first describe how the thick disk considered in my model behaves when irradiated by the white dwarf. Figure \ref{fig:irradiated_disk} shows a diagrammatic representation of the setup. I assume that the stellar rays are parallel, and it can travel inside of the disk till it hits the plane of unit optical depth ($\tau_{*}=1$). For any frequency, I denote the height at which the disk is completely optically thin to a radial ray as:
\begin{equation}\label{eq:height_calc}
    H_{\nu} = h_z\sqrt{\ln\left(\frac{\overline{\sigma}_{~\nu}}{\overline{\sigma}_{ref}}\tau_{0,~\rm ref}\right)}=h_z\chi_{\nu},
\end{equation}
where $\overline{\sigma}_{~\nu}$ is the grain size-weighted extinction cross section at the concerned frequency. As the white dwarf emits mostly in optical/UV, I have $\chi_{*}=\sqrt{\ln(\tau_{0,~\rm ref})}$. The observable I am interested in, however, is mid infrared radiation. Depending on the grain size assumed, $\overline{\sigma}_{\rm MIR}$ can be $\in[0.1,~1]\times\overline{\sigma}_{\rm ref}$, which translates to $\chi_{\rm MIR}\in\sqrt{\ln([0.1,~1]\tau_{0,~\rm ref})}$. For the rest of the section, I assume a fiducial grain size of one micron.

The region where $\tau_{*}<1$, the grains are heated by the white dwarf directly, and, thus, superheated to blackbody temperatures. This region contributes to optically thin dust emission, which I shall discuss in Section \ref{app:optically_thin_cont}. The second emission component is an optically thick emission arising roughly along the locus of $\tau_{*}=1$ (which I term as the disk 'surface'). Here, the superheated dust re-radiates the absorbed starlight and heats up the internal dust grains to similar temperatures through back warming. This occurs mostly near the surface and close to the plane where the incidence angle of the stellar rays onto the surface is close to normal (for example, see figure 3 in \citealt{Chrenko24}). This thus forms a wall of optically thick radiation at high temperatures, which is termed as the radiation from the inner rim. I discuss this component is Section \ref{subsec:inner_rim_ir_flux}. Backwarming can, however, be more efficient and heat up a much larger fraction of the disk. This can lead to enhanced radiation from the disk surface. This is discussed in Section \ref{subsubsec:backwarming}. 

For ease of discussion, I shall focus on the comparison with WD\,1145$+$017 here, and present the calculations for WD\,1232$+$563 in Appendix \ref{app:wd1232}. The results for the latter are broadly similar. However, the primary difference is that its infrared excess is much stronger than that of  WD\,1145$+$017. Thus, a disk with larger $h_z$ (and, correspondingly, lower $\alpha$) is required to explain this system with my model.

\subsubsection{Optically thin dust emission}\label{app:optically_thin_cont}

Significant amounts of optically thin dust heated to high temperatures are known to be present in white dwarf disks. The most compelling evidence comes from the silicate emission features in the mid to far-infrared \citep{Farihi25}. Most past works have indicated that, to match the flux on these emission features, optically thin dust mass of $\sim$~$10^{18\pm1}$~g is required \citep{Ballering22,Farihi25}. Here I primarily focus on the mid-infrared dust continuum emission. The implications of my model on the far-infrared silicate emission feature is briefly discussed in the context of G29-38 in Section~\ref{subsec:g29-38}.

I first compute the total number of grains in this region. Beyond $\chi_{*}h_z$, the entirety of the disk is optically thin to starlight. At lower heights, only a radial fraction of $1/[f(z)\tau_{0,~\rm ref}]$ (which traces the locus of the surface) is optically thin. The total number of dust grains combining both these segments (from both faces of the disk) is then given by (through evaluation of the appropriate integrals):
\begin{equation}\label{eq:nthin} 
\begin{split}
    N_{\rm dust}^{\rm thin}=\frac{4\pi ah_z}{\overline{\sigma}_{\rm ref}}\left[\tau_{0,~\rm ref}{}\int_{\chi_{*}}^{\infty}e^{-x^2}dx + \chi_{*}\right] \\
    + \frac{2\pi h_z\Delta a}{\overline{\sigma}_{\rm ref}}\left[\tau_{0,~\rm ref}\int_{\chi_{*}}^{\infty}e^{-x^2}dx+\frac{1}{\tau_{0,~\rm ref}}\int_{0}^{\chi_{*}}e^{x^2}dx\right].
\end{split}
\end{equation}
In this work, we mostly consider $\Delta a<a$ (motivating from the inference of \citealt{Ballering22}). This makes the $\Delta a$-dependent radial width correction term about an order of magnitude smaller. Thus, for simplicity, henceforth we disregard this term. Inclusion of this term is important when $\Delta a$ approaches $a$. Clearly this term enhances the radiation, thus favoring our discussion.

Assuming my fiducial values of $\tau_{0,~\rm ref}=10^3$, $h_z=R_{\rm WD}$, $a=100~R_{\rm WD}$ (which is consistent with the gray transits, see Figure~\ref{fig:inclination_effect_colorless}), and assuming $\rho_{\rm grain}=3~{\rm g\,cm^{-3}}$ and grain size of one micron, I get a total mass of the optically thin region as $4.5\times10^{17}$~g. Larger assumed particle size yields larger mass: for example, $2.3\times10^{18}$~g with $5~\mu m$ grains. This range of masses is in good agreement with that estimated from observations. Note that, in this configuration, the total mass of the disk is much higher at $\gtrsim$$10^{20}$~g. Thus, thick disks can consistently produce both a large disk mass along with a significant amount of optically thin dust.

However, not all of this dust contribute to the observed radiation. For sufficiently edge-off views (or larger inclinations), one face of the disk is obscured by the optically thick disk midplane. For edge-on views, though part of both faces are visible, there is significant self shielding from the front half of the disk. These reduce the amount of radiation reaching the observer significantly. For my purpose, I adopt two crude (and rather conservative) corrections to Equation \ref{eq:nthin}: 1) I consider only half of the dust to account for the self-shielding and, 2) for the second term inside the parenthesis, I use $\chi_{*}-\chi_{\rm MIR}$ instead to consider only the portion of the disk radially optically thin to mid infrared.\footnote{For the $\Delta a$ term, the correction would be to change the lower limit of the second integral inside paranthesis to $\chi_{\rm MIR}$. But this term is anyway neglected in this work.} I denote this revised estimate as $N_{\rm dust}^{\rm thin,~MIR}$. The associated mass of the dust then reduces to $2\times10^{16}$~g (with one micron grains). 

As noted previously, the dust in this region is heated directly by the star light. The dust temperature, thus, is given by:
\begin{equation}\label{eq:trim}
    T_s\approx \left[\frac{\overline{\sigma}^{\rm abs}_{P}(T_{\rm WD})}{\overline{\sigma}^{\rm abs}_{P}(T_{\rm rim})}\right]^{1/4}\left(\frac{R_{\rm WD}}{na}\right)^{1/2}T_{\rm WD}.
\end{equation}
where $n=2~(1)$ when the disk is optically thin (thick) to infrared dust emission, and $\overline{\sigma}^{\rm abs}_{P}(T)$ is the Planck-function weighted absorption cross section at temperature $T$. The solution of this equation requires an iterative process. Manual inspection shows that, for WD\,1145$+$017, this results in a temperature roughly in the range of $1500\pm500$~K for $a=100~R_{\rm WD}$. Note that the observed blackbody temperature of the infrared flux in \cite{Vanderburg15} is $\approx1145$~K, well within this regime.\footnote{Note that here the infrared temperature and the transit period indicates similar disk radius. This hints that the dust and debris disks may well be related.} Thus, I use this as my $T_{\rm rim}$ to check if it yields the desired flux.

I focus on the WISE W1 $3.4~\mu m$ and W2 $4.6~\mu m$ fluxes for comparison. The total flux for WD\,1145$+$017 is then given by (assuming my fiducial values above):
\begin{equation}\label{eq:othin_flux}
\begin{split}
    F_{\rm MIR} = \frac{N_{\rm dust}^{\rm thin,~MIR}~\overline{\sigma}_{\rm MIR}^{\rm em}}{d^2}B_{\nu}(T_s)\\ 
    \approx2\times10^{-3}~{\rm mJy}\left(\frac{a}{100~R_{\rm WD}}\right)\left(\frac{h_z}{R_{\rm WD}}\right)\left(\frac{\overline{\sigma}_{\rm MIR}^{\rm em}/\overline{\sigma}_{\rm ref}}{0.1}\right),
\end{split}
\end{equation}
where $\overline{\sigma}_{\rm X}^{\rm em}$ is the particle distribution weighted emission (same as absorption) cross section at wavelength $\rm X$. 

The measured W1 and W2 flux for WD\,1145$+$017 are $(5.04\pm0.56)\times10^{-2}$~mJy and $(4.40\pm1.11)\times10^{-2}$~mJy, respectively \citep{Morocco21}. But $\approx0.02~(0.01)$~mJy is contributed by the star in the W1 (W2) band \citep{Vanderburg15}. Thus the disk flux is $\approx(3.04\pm0.56)\times10^{-2}$~mJy and $(3.40\pm1.11)\times10^{-2}$~mJy in W1 and W2, respectively. Compared to this, the flux calculated from the optically thin dust is about an order of magnitude lower. This indicates that this component may not be sufficient to explain the mid-infrared excess. The contribution, however, is non-negligible. Additionally, the flux is strongly dependent on the disk ($a$, $h_z$) and dust parameters. Altering them may result in a significant increment of the flux. I arrive at similar conclusion for WD\,1232$+$563 (see Appendix \ref{app:wd1232})

\subsubsection{Flux from the Inner Rim}\label{subsec:inner_rim_ir_flux}

This radiation component has been extensively modeled in the context of protoplanetary disks, and often invoked to explain the near to mid-infrared bumps seen in their spectral energy distributions (see \citealt{Dullemond10} for a great and concise review on this topic). It has, however, been not much explored in the context of white dwarf disks. Recently though, the presence of this radiation component has been noted in the numerical simulations by \cite{Ballering22} in the context of G29-38. Proper modeling of this radiation is very complex and requires appropriate consideration of the rim geometry, temperature variation along the surface, and projection effects (see for example \citealt{Isella05}). For my purpose, I employ the simple model proposed in \cite{Dullemond03}.

They model the inner rim as a vertical isothermal cylindrical blackbody wall. Assigning an appropriate height, $H_{\rm rim}$ to this wall is a challenge. \cite{Dullemond03} prescribes this height to be (in my notation) $\chi_{*}h_z$. However, detailed modeling and numerical simulations suggests that this may not be a proper assumption as the $\tau_{*}=1$ surface is clearly not vertical (\citealt{Isella05,Flock16,Chrenko24}, and also Figure \ref{fig:irradiated_disk}). For my purpose, I use a conservative value of $H_{\rm rim}=h_z$. They then define:
\begin{equation}
    \delta = \frac{H_{\rm rim}}{R_{\rm rim}}\tan(i)
\end{equation}
where $R_{\rm rim}$ is the distance to the inner rim from the white dwarf. For WD\,1145$+$017, I take $R_{\rm rim}=a=100~R_{\rm WD}$. The flux as a function of inclination angle $i$ is then analytically given by:
\begin{equation}\label{eq:inner_edge_flux}
    F_{\nu}=2B_{\nu}(T_{\rm rim})\left(\frac{R_{\rm rim}}{d}\right)^2\cos(i)\left(\delta\sqrt{1-\delta^2}+\arcsin(\delta)\right)
\end{equation}
when $\delta<1$ and
\begin{equation}
    F_{\nu}=\pi B_{\nu}(T_{\rm rim})\left(\frac{R_{\rm rim}}{d}\right)^2\cos(i)
\end{equation}
when $\delta>1$. Here, $d$ is the distance to the white dwarf and $T_{\rm rim}$ is the temperature of the rim. This is approximately given by the same expression as $T_s$ (Equation \ref{eq:trim}, \citealt{Dullemond03}). Thus, as in the previous section, I set $T_{\rm rim}=1145$~K for WD\,1145$+$017. The effect of this parameter (and $R_{\rm rim}$) is briefly discussed in Appendix~\ref{app:param_study}.

\begin{figure}[t]
    \centering
    \includegraphics[width=1.0\linewidth]{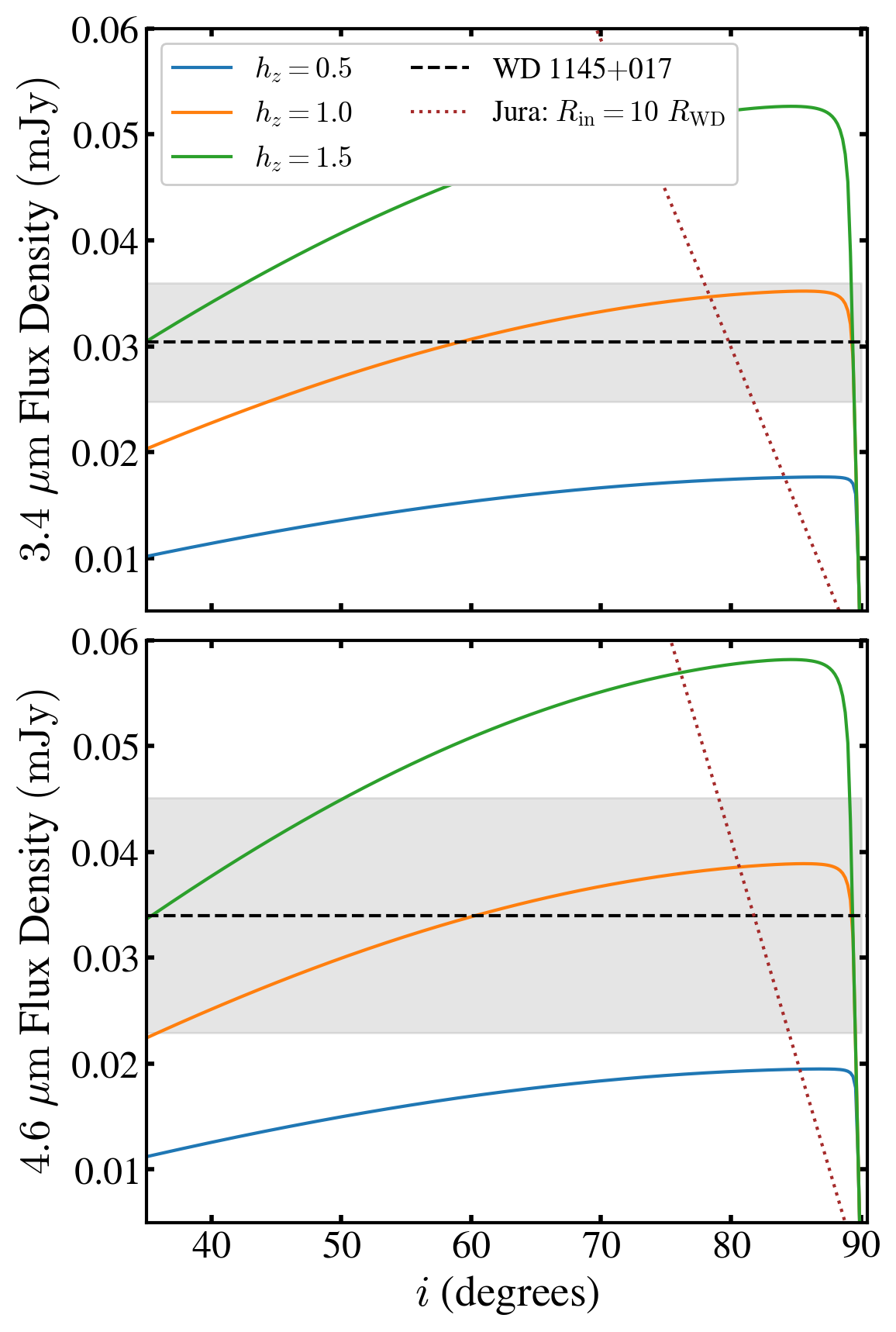}
    \caption{The predicted mid-infrared WISE W1 $(3.4~\rm \mu m)$ and W2 $(4.6~\rm \mu m)$ fluxes from the inner rim (at $R_{\rm rim}=100~R_{\rm WD}$) as a function of inclination for different $H_{\rm rim}=h_z$. The horizontal line denotes the measured excess flux for WD\,1145$+$017. It is seen that the inner rim can produce sufficient mid infrared flux to explain this excess even at very high inclinations consistent with the transits. For comparison, I plot the flux from a much compact (inner radius at $10~R_{\rm WD}$) flat disk \citep{Jura03} model. This requires $i\approx80$~degrees to meet the required flux, inconsistent with the transits.}
    \label{fig:inner_edge_flux}
\end{figure}

The resultant flux as a function of $i$ is shown in Figure \ref{fig:inner_edge_flux}. The first intriguing aspect is that, unlike any other component, the flux peaks at a very high inclination angle of $\gtrsim$$85$ degrees (though this might get modified with detailed modeling of the inner rim shape and its projection, see \citealt{Isella05}). Post this the flux drops rapidly, but given the high temperature, it is still bright. Comparing to the WISE fluxes of WD\,1145$+$017, I find that it is indeed possible for this component to yield the entire observed infrared excess flux at very high (but edge off) inclination angles. I obtain similar results for WD\,1232$+$563 in Appendix~\ref{app:wd1232}, albeit with a larger $h_z$ (and correspondingly lower $\alpha$ for maintaining colorlessness in transits). This establishes this radiation component to be an important and potentially large contributor to the near to mid infrared excess of white dwarfs and cannot be ignored.   

\subsubsection{Efficient Backwarming and optically thick emission}\label{subsubsec:backwarming}

In the formalism of \cite{Chiang97}, it is expected that deep inside the disk, the temperature is given by Equation \ref{eq:trim}. For my fiducial disk parameters for WD\,1145$+$017, this evaluates to a temperature of a few hundred kelvins. At such low temperatures, the emission would be sub-dominant compared to the effects discussed above. However, the temperature distribution found by \cite{Ballering22} in their radiative transfer simulation for G29-38 (where the disk radius is also $\approx$$100~R_{\rm WD}$) does not agree with this picture. Referring to their figure 5, almost all of the disk appears to have significantly high temperatures. A dearth of cold dust is also observationally indicated through lack of significant far-infrared emission \citep{Farihi16}.\footnote{Most of the measurements, however, provide only upper limits on the $24~\rm \mu m$ flux, see \citealt{Jura07}. Also, cold dust continuum peaking at mid infrared can be hard to detect due to dominance by the optically thin dust emission. This motivates deeper far-infrared observations of white dwarfs to detect (or rule out) cold gas.}

This may result from backwarming in the disk (see \citealt{Calvet91} for the analytic solution in plane-parallel atmosphere, also figure 1 in \citealt{Kama09}), where the heat from the super-heated surface dust efficiently cascades inwards, through successive absorption and re-radiation of the infrared photons,  to heat up the inner disk. This will have a profound effect on the spectral energy distribution, as, even disregarding the effects discussed in the previous sections, it will greatly enhance the optically thick disk surface continuum.\footnote{This is different from the classical flat disk model as here the disk can be placed much further away without compromising with the temperature. This greatly enhances the emission surface area.} Additionally, it will also result in an optically thick `outer rim' of similar height as the inner rim. Assuming the limiting case that the entire disk is heated almost uniformly to $T_{\rm disk}$, I get the total flux from the disk surface and outer rim to be:
\begin{equation}
    F_{\nu}\approx\frac{2a}{d^2}B_{\nu}(T_{\rm disk})\left[\pi\Delta a\cos(i)+2h_z\sin(i)\right],
\end{equation} 
where the second term comes from the projection of the outer rim. The first term dominates almost for all $i$, except at the edge-off cases where the second term survives. Inspection shows that $T_{\rm dust}\gtrsim1000$~K is needed to satisfy the excess in WD\,1145$+$017 at inclinations consistent with edge-off transits. Though not impossible, such a high temperature throughout the disk is unlikely. However, the contribution to the mid infrared flux can still be significant (definitely so for low inclination angles) even with lower temperatures. For instance, with an overall disk temperature of $\approx800$~K (referring to \citealt{Ballering22}) yields an edge-off flux from the outer rim of $\approx0.01$~mJy. This is significant and takes away burden from the inner rim to meet the required flux alone. I arrive at similar conclusions for WD\,1232$+$563 in Appendix~\ref{app:wd1232}.

\subsection{Reconciling with Transit Observations}\label{subsec:reconciling_with_transits}

The thick-disk radiation components discussed above allow detection of infrared excess along with transits without requiring disk misalignment. In my model, the dominant effect in the near to mid infrared is the emission from the inner rim, thus I focus on this component. In case of WD\,1145$+$017, inspection shows that $i$$\lesssim$$89.5$ degrees is required for the inner rim to explain the observed flux. This is well within the range of angles that allow transits (unlike with the flat disk model, which requires $i$$\lesssim$$80.0$: inconsistent with transit observations with an aligned debris disk). The exact allowed range, however, depends on assumed $h_z$. The case of $h_z$$\approx$$1$ is interesting as it yields the required flux over a broad range of edge-off angles, which makes it consistent with the range of angles given by Figure~\ref{fig:inclination_effect_colorless} for gray transits. With larger $h_z$, however, the constraint from infrared flux is tighter at $89$$<$$i$$<$$89.5$ degrees. But that would yield a much larger transit depth than observed with the assumed optical depth, thus unlikely. We arrive at a similar conclusion for WD\,1232$+$563, shown in Appendix~\ref{app:wd1232}. It is also seen from Figure~\ref{fig:inner_edge_flux} that the infrared flux from the rim is a strong function of $h_z$. Thus, absence of infrared excess in any transiting system can also be explained with a small disk height.

I now briefly revisit the unique case of WD\,1013$-$0427. The system was not detected in WISE even during the transit event. The derived upper limits in both W1 and W2 are $\approx$$0.025$~mJy. Reconciling this with the inferences in Section~\ref{subsubsec:wd1013}, which suggests a high disk height ($h_z/R_{\rm WD}\approx1.75$), is, however, not straightforward. In most configurations, my model predicts the inner rim radiation to be bright enough to be detectable. The WISE limits are only approximately met when the disk is placed farther away at $\approx$$300~R_{\rm WD}$ (against the assumed $a=100~R_{\rm WD}$ in Section~\ref{subsubsec:wd1013}) and a corresponding temperature of $T_{\rm rim}\approx1000$~K is assumed. Such a disk radius significantly limits the range of edge-off angles for witnessing transits. But the effects on the depth and color remain identical. See Appendix~\ref{app:transit_a_dep} for the transit calculations for this configuration. But it remains to be checked if such a configuration is physically consistent.

\subsection{Application to low inclination system: G29-38}\label{subsec:g29-38}

Here, I briefly discuss the general applicability of my thick-disk model to low inclination systems by considering the case of G29-38. The inclination of this system has been derived to be 30 degrees in \citealt{Xu18irvar}. Note that this derivation uses the flat disk model which may be revised under thick disk assumption. For my purpose, like in \cite{Ballering22}, I continue with $i=30$~degrees. The system shows significant infrared excess, including the silicate emission feature at 10~$\rm \mu m$. This is, thus, a great test case for both the optically thick and thin radiation components of the disk. 

I take the SED of G\,29$-$38 (from table~1 in \citealt{Ballering22}) and the Spitzer IRS spectrum (\citealt{Reach09}, kindly shared with me by William T. Reach). I consider all the three thick disk emission components discussed above. All the formulations remain the same, except we find $\chi_{\rm IR}$ such that
\begin{equation}\label{eq:ch_ir_incl}
    \frac{\overline{\sigma}_{~\nu}}{\overline{\sigma}_{ref}}\frac{\tau_{0,~\rm ref}}{\Delta a}\int_{\chi_{\rm IR}h_z}^{\chi_{\rm IR}h_z+\Delta a\cot(i)}f(x)\frac{dx}{\cos(i)}=1,
\end{equation}
and use $\chi_*-\chi_{\rm IR}$ in Equation~\ref{eq:nthin} to get the total number of dust particles emitting in optically thin regime.\footnote{It is unclear what infrared cross section to use to determine the optically thin regime, as part of scattered light is also received by the observer. I am currently using extinction cross section, but manual inspection shows that using absorption or effective $\left(\sqrt{\sigma^{\rm abs}(\sigma^{\rm abs}+\sigma^{\rm sca})}\right)$ cross section do not alter the conclusions and results only mild quantitative changes.} For simplicity, I assume a single grain size of one micron and the grain composition to be that of astronomical silicate. The refractive indices were taken from \cite{Draine03b}.\footnote{Downloaded from \url{https://www.astro.princeton.edu/~draine/dust/diel/callindex.out_silD03}. Note that this is not strictly correct as these values are for 0.1 micron grains instead. But comparison of the absorption cross section with figure 24.1 in \cite{Draine11} for one micron grain show no significant differences.} The cross sections were calculated using the \texttt{Python} package \texttt{miepython} \citep{miepython23}. I then solve Equation~\ref{eq:ch_ir_incl} to get the $\chi_{\rm IR}$ as function of wavelength. For the rest of the parameters I employ trial-and-error (mostly within the inference range in \citealt{Ballering22}) to compare the model and the observed SED.

\begin{figure}
    \centering
    \includegraphics[width=1\linewidth]{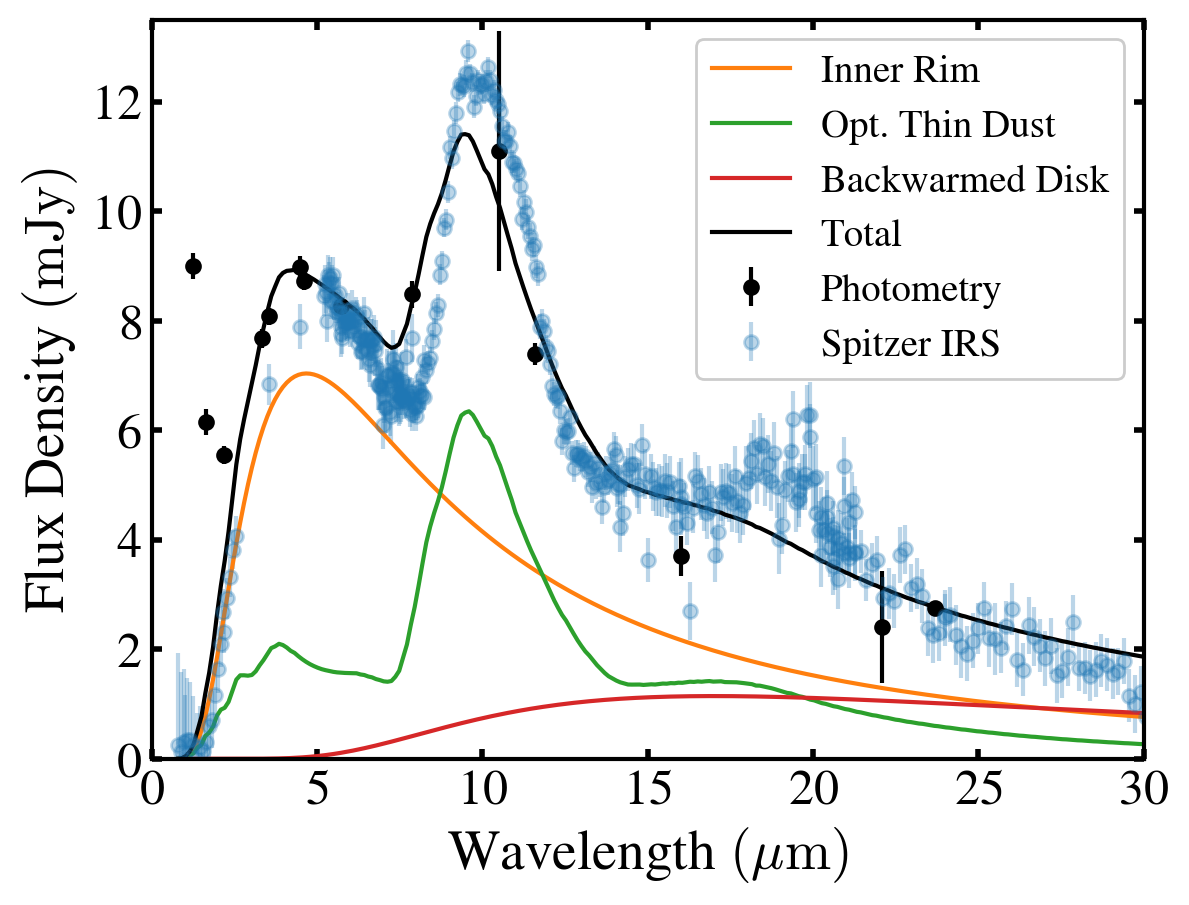}
    \caption{A representative visual fit of my infrared emission model to the SED of G29-38. The total flux as well as the separate components are shown. The model can capture all the essential SED features including the silicate emission feature. The first few data points are from the white dwarf.}
    \label{fig:G2938_sed}
\end{figure}

I find that my model can capture most of the features in the SED (including the silicate emission) with a range of parameters. One of the visual best fits is shown in Figure \ref{fig:G2938_sed}. This corresponds to $a=100~R_{\rm WD}$, $\Delta a=15~R_{\rm WD}$ and a relatively large disk height of $h_z=6.0~R_{\rm WD}$ (corresponding to an opening angle of $\approx$$3.4$ degrees). The inner rim and optically thin dust temperature has been set to $1090$~K. These are all broadly consistent with the constraints of \cite{Ballering22}, except the $\Delta a$ being higher in my model. This is required to attain the large silicate emission feature (see Appendix~\ref{app:si_em_par_study} for how the disk parameters affect the silicate emission feature). Figure~\ref{fig:G2938_sed} also compares well with figure~1 in \cite{Ballering22} and figures~5 and 6 in \cite{Xu18irvar}. The inner disk temperature, however, needs to be much lower in my model ($T_{\rm disk}$$\lesssim$$400$~K), which indicates weak backwarming. However, owing to (visual) degeneracies similar fits can be achieved with different model parameters. For example, reduction in $\Delta a$ would need increase in $h_z$, but can accommodate higher $T_{\rm disk}$. My model, however, cannot adequately capture the shape of the emission feature together (similar difficulty noted in \citealt{Ballering22}), and the bump around $20~\rm \mu m$. I suspect the reason behind this to be the inadequacy of having only astronomical silicate grains. It it thus important to model the emission features with different types of grains and constrain the composition (for example, see \citealt{Reach09}).

One primary difference, however, lies in the total mass of the disk. \cite{Ballering22} infers a total mass of $\approx$$5\times10^{18}$~g. This is similar to the total optically thin dust mass in my model which is $\approx$$2.5\times10^{18}$~g (though the mass from which the radiation is received is about a factor of eight lower due to the corrections discussed in Section \ref{app:optically_thin_cont}). The total mass of the disk, however, is much larger at $\approx8\times10^{20}$~g. This clearly shows that a thick disk can hold a significant amount of dust within its optically thick core. This can potentially address the higher amounts of accreted mass onto white dwarfs than estimated solely from the optically thin disk component. 

\subsection{Implications and Future Studies}

It is clear that there are several complex effects at play in thick disks which can significantly affect the disk infrared emission. It would thus be of interest to explore the relative importance of these effects in future studies. Extensive numerical simulations of disks around white dwarf, similar to \cite{Ballering22} need to be performed. 

I now discuss a few observational probes. One of them is mid infrared variability. Infrared variability has been detected in several dusty white dwarfs \citep{Xu14,Xu18irvar,Swan19,Rogers20,Guidry24b}. This might speak in favor of emission from optically thin dust (proportional to the amount of dust, thus susceptible to variability) as the dominant contributor. However, the two transiting debris systems with infrared excess do not show any mid infrared variability \citep{Rogers20,Hermes25}, even though significant photometric variability exists. Lack of variability is more consistent with emission from the optically thick components. 

The second probe is the dust emission features, which can constrain the amount of optically thin dust well. Recently, JWST has been instrumental in detecting spectacular emissions from several dusty white dwarfs \cite{Farihi25}. Similar study is needed for transiting debris systems, where the viewing inclination is known to be high. This would give novel constraints in the structure of the optically thin dust in the disk.

\section{Summary, Limitations and Conclusions} \label{sec:conclusions}

A significant fraction of white dwarfs host dust\slash debris disks formed from the tidal disruption of asteroids and planetesimals. Two primary probes of such disks are 1) excess infrared emission from the heated dust, and 2) photometric transits by the disk when viewed at high inclinations. Past canonical models mostly use either purely optically thick and geometrically thin (a.k.a flat) disk or a purely diffused optically thin cloud. These models, though have been widely successful in explaining many white dwarf systems, face certain limitations with regards to several recent observations, particularly for high inclination systems. For example, the infrared flux from a flat disk falls short of that observed for two transiting systems WD\,1145$+$017 and WD\,1232$+$563. More recently, the reddening observed in the long transit in WD\,1013$-$0427 cannot be satisfactorily explained with the flat disk model. Additionally, a pure flat disk cannot produce optically thin dust emission features seen in many white dwarfs. A diffuse optically thin cloud, on the other hand, can explain some of these observations. It's limitation, however, is the inability to contain enough mass (and still remain diffuse) to explain the amount of material accreted onto white dwarfs.

In this work, I instead consider a geometrically thick disk and explore the effect on the above observables. In particular, I assume a disk with a Gaussian vertical profile with scale heights comparable to the white dwarf radius. Such a disk has optically thin outer layers and optically thick inner layers. I investigate how the various disk properties, namely scale height, $h_z$, inclination, $i$ and mid-plane optical depth, $\tau_{0,~\rm ref}$ affect the transit as well as infrared observations. I primarily focus on the high inclination regime which yields transits, but discuss generalizations to other systems whenever possible.

I first study transits in the case of pure edge-on view ($i=90$~degrees, Section \ref{subsec:edgeon}) to show that for a given $\tau_{0,~\rm ref}$, there is a limit to the maximum color that can be achieved by changing $h_z$. Comparing with WD\,J1013$-$0427, I confirm the inferences of \cite{Bhattacharjee25} that the occulting dust is required to not be optically thick and abundantly populated with small grains (resulting in an angstrom exponent $\alpha\approx3$) to explain the observed transit color.

Following this, I go beyond the perfect edge-on view (which is statistically improbable with random orientation angles) to include edge-off viewing angles (Section \ref{sec:edgeoff}). I show that thick disks allow for significant transits even at edge-off inclinations. I also show that, owing to the outer layers of the disk being less dense, edge-off viewing can induce significant transit color in an otherwise optically thick disk. I show that the transit properties of WD\,J1013$-$0427 can indeed be satisfied with edge-off views of disks with high mid-plane optical depths. With regards to the gray transitors, I show that it is possible for the transits to be gray at edge-off inclinations when $\alpha$$\lesssim$$0.5$. This, however, does not imply a complete lack of small particles (see Section \ref{app:angstrom_exp}).

The primary limitation of the above analysis is the non-inclusion of larger particles, like asteroid-simals, in my model. These large bodies can contribute significantly to the transits, thus reducing the burden on the dust alone to meet the observed transit depths. This nothing but aids my discussion, especially with the gray systems. Additionally this model does not explain in itself why different disks are expected to exhibit very different $\alpha$ values.

I then proceed to examine the effect of the disk thickness on the infrared emission. I mainly consider the radiation components specific to thick disks (and lacking in thin disks) and estimate the resulting flux for the high inclination (transiting) systems. Specifically, I consider three radiation components: 1) emission from the outer optically thin dust (Section \ref{app:optically_thin_cont}), 2) emission from the superheated disk inner rim (Section \ref{subsec:inner_rim_ir_flux}), and 3) optically thick emission from disk interior heated through efficient backwarming (Section \ref{subsubsec:backwarming}). I show that, even with moderately thick disks, all these effects can significantly contribute to the infrared radiation at all inclinations. These radiation components (especially the flux from the inner rim) are enough to explain the observed infrared excess in WD\,1145$+$027 and WD\,1232$+$563, consistently with the transits. Additionally, the model can also produce the SED of the low inclination system G29-38 including the silicate emission feature in the far infrared, demonstrating its general applicability (Section \ref{subsec:g29-38}). These show that thick disk effects cannot be neglected when fitting the infrared emission from the dust disk, which, otherwise, may lead to inaccuracies in the inferred disk geometry. 

The limitation in this aspect is the simplicity of the calculations, as, in reality, the above effects are much more complex. A proper treatment would require numerical radiative transfer simulations of thick disks, similar to as performed in \cite{Ballering22}. Such a detailed study is beyond the scope this work, but it's need is established.

Overall, this work demonstrates that even moderately thick disks can have profound effects on the transit and infrared properties of disk-hosting white dwarf systems. Given that several studies indicate that the white dwarf disks can indeed be thick (primarily due to collisional cascade), it would be important to consider and study the associated effects. More theoretical as well as observational studies need to be performed. Detailed radiative transfer simulations would help quantify the importance and contributions of these effects. With regards to observations, I think constraining the infrared excess in high inclination (transiting) systems, and possibly detect silicate emission features therein with new instruments like JWST, would provide significant insights into the disk structure. Infrared surveys like SPHEREx will also provide data crucial to understand the dust and debris disks in these systems.

\section{Acknowledgements}

I thank the anonymous referee whose comments improved the content and clarity of the paper significantly. I am grateful to Siyi Xu and Lynne Hillenbrand for very helpful discussions and suggestions throughout, reading the draft and providing detailed comments. I also want to thank Adolfo Carvalho for a great pedagogical discussion on protoplanetary disks which helped me understand the various radiation effects discussed in the paper. I also thank Nick Ballering for ascertaining the result from his simulations of the inner edge flux potentially contributing significantly to the infrared flux. Finally, I thank Nick Ballering, Shri Kulkarni, JJ Hermes, Joseph Guidry, Amy Bonsor, Yuqi Li, and William T. Reach for reading the draft and providing very helpful comments.

I have used \texttt{Python} packages Numpy \citep{harris2020array}, SciPy \citep{2020SciPy-NMeth}, Matplotlib \citep{Hunter:2007}, Pandas \citep{reback2020pandas}, Astropy \citep{Astropy13, Astropy18}, and Astroquery \citep{astroquery19} at various stages of this research.

\section{Data Availability}

All the relevant codes to reproduce the results of this paper can be found in \url{https://github.com/Soumin1908/wd_thick_disk_models}

\appendix

\section{Angstrom Exponent}\label{app:angstrom_exp}

\begin{figure*}[t]
    \centering
    \includegraphics[width=1.0\linewidth]{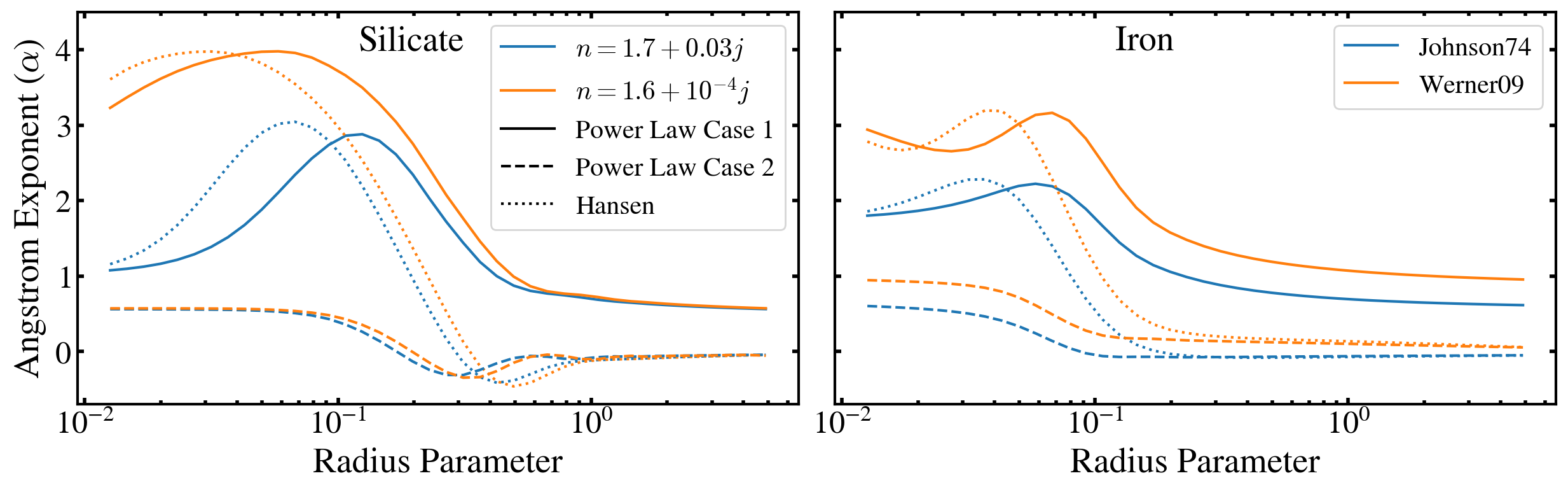}
    \caption{The effective Angstrom exponent, $\alpha$, as a function of grain size parameters refractory silicates (left panel) and pure iron (right panel). I consider two different refractive indices, $n$. For each material I consider three grain size distributions: power-law with fixed $r_{\rm min}$ at $0.01~\rm \mu m$ ($r_{\rm max}$ is parameter varied, Power Law Case 1, solid line), power-law with fixed $r_{\rm max}$ at $5~\rm \mu m$ ($r_{\rm min}$ is varied, Power Law Case 2, dashed line) and Hansen distribution ($r_{\rm eff}$ is varied, dotted line). It is seen that the possible values of $\alpha$ depends significantly on the size distribution and the material. However, in any reasonable parameter space, it is difficult to achieve $\alpha>3$.}
    \label{fig:angstrom_exponent}
\end{figure*}

Here I investigate how the Angstrom exponent $\alpha$ depends on the grain size distribution and particle type. Because the majority of the transiting objects are observed in the optical light curve surveys, I consider the optical wavelengths. I evaluate Equation \ref{eq:sigma_mean} for a few different cases over the range of optical wavelength ($0.35\mu m$ to $0.9\mu m$) and fit with a straight line in the log space using the \texttt{curve\_fit} module in \texttt{scipy}. The extinction cross sections have been computed using the python package \texttt{miepython} \citep{miepython23}. As in \cite{Bhattacharjee25}, I consider two particle distributions. First is the power law given by 
\begin{equation}\label{eq:pl_dist}
    n(r)\propto r^{-p},
\end{equation}
where I assume $p=3.5$, which is expected from collisional cascade and that inferred in other setups \citep[interstellar medium, see][]{Mathis77}. Following \cite{Croll14}, I fix the minimum radius $r_{\rm min}$ to $0.01~\mu m$ and vary the maximum radius, $r_{\rm max}$ (Power Law Case 1). However, I note that other works like \cite{Ballering22} fixes $r_{\rm max}$ to a reasonably large value of $5~\mu m$ to vary $r_{\rm min}$. Thus, I also consider this case (Power Law Case 2, and show that the two choices yield drastically different results).

The second is from \citet[][Hansen distribution]{Hansen71}, which, in context of debris transits, has been used in \citet{Hallakoun17}:
\begin{equation}
    n(r)\propto r^{\frac{1-3\nu_{\rm eff}}{\nu_{\rm eff}}}e^{-\frac{r}{\nu_{\rm eff}r_{\rm eff}}}
\end{equation}
where $\nu_{\rm eff}$ and $r_{\rm eff}$ are the effective variance and radius, respectively. Following \citet{Hallakoun17}, I use $\nu_{\rm eff}=0.1\ \mu$m, $r_{\rm min}=0$ and $r_{\rm max}=\infty$ as fixed parameters. The variable in this model is $r_{\rm eff}$.

For each case of particle distribution, I consider several different kinds of dust materials, characterized by different complex refractive indices. The first group of material are the refractory silicates. I consider the astronomical silicate as in \cite{Hallakoun17} ($n=1.7+0.03j$), the refractive index being quite stable over the whole optical wavelengths. Note that this refractive index is very close to that of Corundum \citep{Croll14}. I also consider the case of enstatite ($n=1.6+10^{-3.5}j$) as a case with very low absorptive component. The second material is pure iron, as considered in \cite{Croll14}. But this is trickier as the refractive index of iron changes significantly over the optical wavelength range, making it suboptimal to assume one single value (though several works have done that in the past, including \citealt{Bhattacharjee25}). To be more correct, I download the data from \url{https://refractiveindex.info/}. I notice that two studies (Johnson74 and Werner09) seem to show very different results. Thus, I consider both the cases. 

The results are shown in Figure \ref{fig:angstrom_exponent}. Firstly, I see that it is quite difficult to achieve $\alpha\gtrsim3$ with any reasonable grain property and distribution. Silicates with very low absorptive component is capable of achieving high $\alpha$, however such grains are not expected to be the dominant component in disks. Another interesting aspect is that the Hansen and Power Law Case 2 distributions can yield $\alpha<0$ and, thus, can potentially induce bluing in the transits: a possibility explored for WD\,1145$+$017 in \cite{Hallakoun17} with the Hansen distribution. Finally, I note that all the three distributions yield very different results. Choosing one over the other for any simulation can, thus, lead to significantly different outcomes. Thus, it is of utmost importance to either justify a choice, or span a broad range of distribution and make conclusions accordingly.

\section{Effect of disk radius on edge-off transit}\label{app:transit_a_dep}

\begin{figure}[t]
    \centering
    \includegraphics[width=1.0\linewidth]{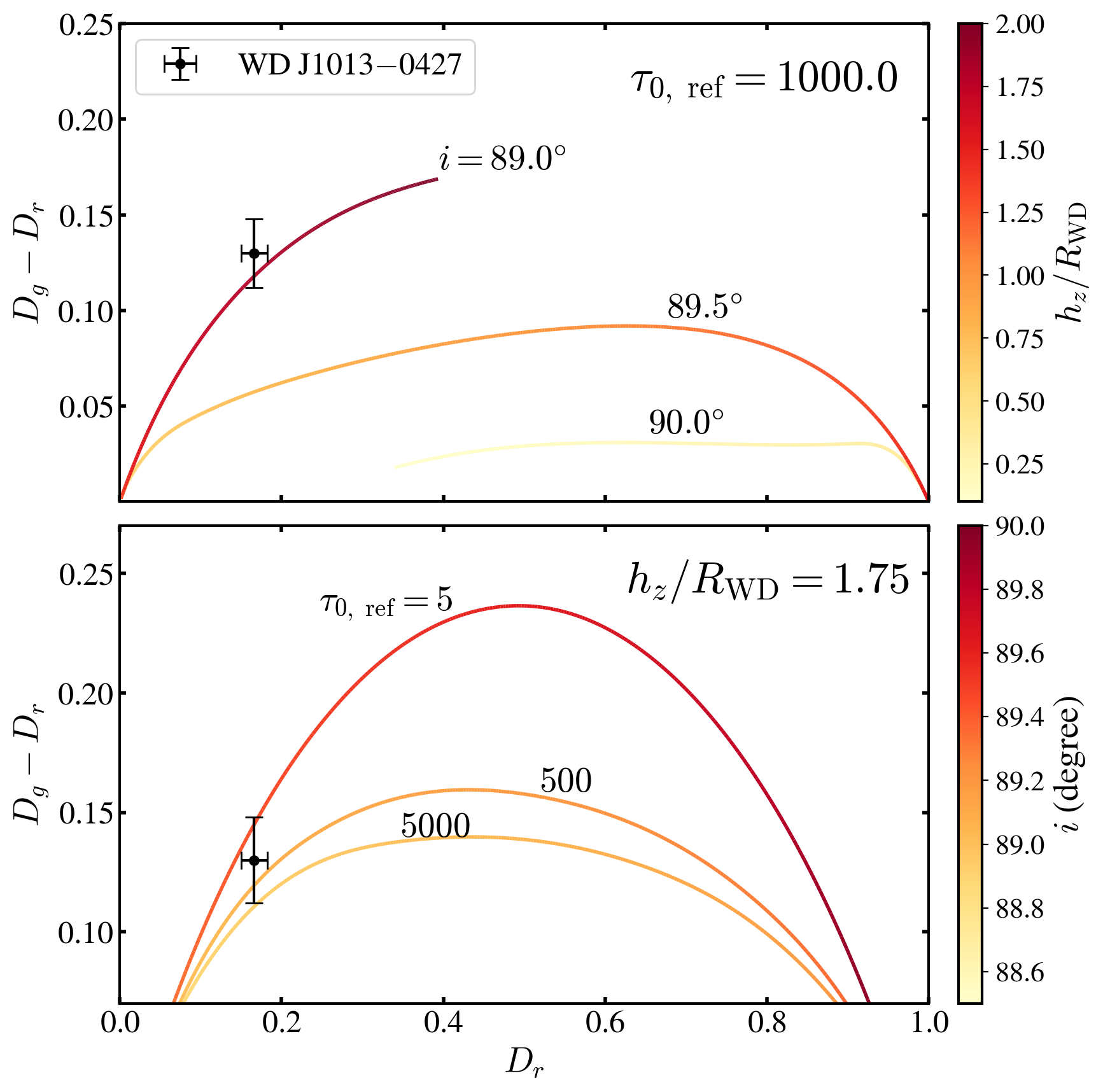}
    \caption{Same as Figure \ref{fig:inclination_effect} but with $a=300~R_{\rm WD}$ instead. The shape of the curves in the depth-color space is qualitatively unchanged. The range of allowed angles, however, has decreased significantly.}
    \label{fig:inclination_effect_a300}
\end{figure}

In Section \ref{subsubsec:wd1013} I assumed a fiducial case of $a=100~R_{\rm WD}$ for WD\,J1013$-$0427. However, as discussed in Section~\ref{subsec:reconciling_with_transits}, the non-detection in WISE suggests a wider orbit configuration. Here, I show the same figure as Figure \ref{fig:inclination_effect} but with $a=300~R_{\rm WD}$. I find that there is no qualitative change in the results and all the discussions in Section \ref{subsubsec:wd1013} still holds. The only difference is that the range of viewing angles that can yield the transit and color has reduced drastically. This is easy to understand. At larger distance, a small change in inclination will result in a large vertical shift within the disk (see Equation \ref{eq:projection}), where the optical depth would be much lower. However, larger disk scale heights, than the range assumed in this work, can increase the range of possible viewing angles. This maybe applicable to larger and more vigorous collision events at wide orbits, a possibility for WD\,J1013$-$0427 discussed in \cite{Bhattacharjee25}.

Changing $a$ has identical effect on the gray transitors. To avoid redundancy I do not discuss it here, as the assumed $a$ for WD\,1232$+$563 (the next section) is larger, and the effect is automatically seen. 

\section{WD\,1232+563}\label{app:wd1232}

This is the only published object other than WD\,1145$+$017 to show both photometric transits \citep{Guidry21,Hermes25} and infrared excess \cite{Debes11}. It has a temperature of $T_{\rm WD}=11,787$~K and is at a distance of $171.92$~pc. The WISE W1 and W2 fluxes (after approximately subtracting the stellar contribution) are $\approx0.052\pm0.003$~mJy and $\approx0.068\pm0.007$~mJy, respectively. \cite{Debes11} inferred a inner dust disk radius of $7~R_{\rm WD}$ using the classical flat disk which corresponds to an inner disk temperature of $1846$~K. The WISE colors, however, indicate a much lower effective temperature of $\approx$$1000$~K. Using Equation~\ref{eq:trim}, this yields a distance to the disk of $\approx$$150~R_{\rm WD}$, much further away than given by the classical model. With regards to transits, unlike WD\,1145$+$017, there is no confident period detection. But \cite{Hermes25} detected a possible period at $14.8$~hours, which places the transit-causing disk at $\approx$$200~R_{\rm WD}$ (interestingly, close to the revised distance of the dust disk). I use these respective distances in my calculations.

The infrared continuum from the optically thin dust, with the assumed dust temperature and distance yields:
\begin{equation}
    F_{\rm MIR} \approx1\times10^{-3}~{\rm mJy}\left(\frac{a}{150~R_{\rm WD}}\right)\left(\frac{h_z}{R_{\rm WD}}\right)\left(\frac{\overline{\sigma}_{\rm NIR}^{\rm em}/\overline{\sigma}_{\rm ref}}{0.1}\right).
\end{equation}
This is more than an order of magnitude less than the observed flux. Though larger $h_z$ can increase the flux proportionately, it is unlikely that it will be able to satisfy the entire observed excess.

\begin{figure}
    \centering
    \includegraphics[width=1\linewidth]{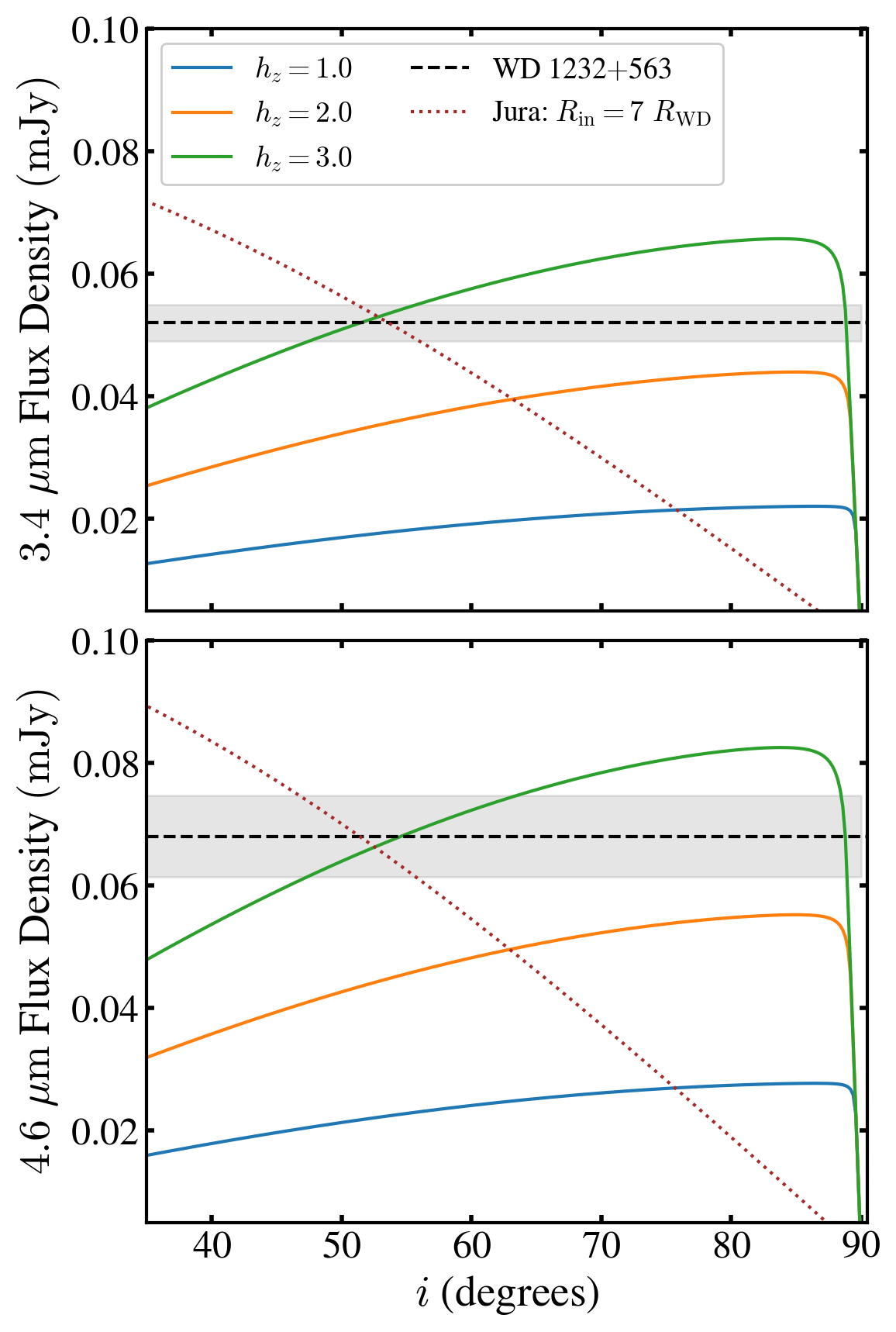}
    \caption{Flux from the inner rim for WD\,1232$+$563. Same format as Figure \ref{fig:inner_edge_flux}, but with $a=150~R_{\rm WD}$ and $T_{\rm rim}=1000$~K instead. I find that, even in this case, the flux from the inner edge at edge-off angles is sufficient to explain the mid infrared excess, albeit requiring the disk to be much thicker. For comparison, I include the flat model by \cite{Jura03} for inner radius of $7~R_{\rm WD}$, as derived in \cite{Debes11}. The required inclination in this case is very high at $\approx$$55$~degrees (consistent with the inference in \citealt{Debes11}). But such a large inclination cannot produce transits in aligned disks.}
    \label{fig:inner_edge_flux_wd1232}
\end{figure}

\begin{figure}
    \centering
    \includegraphics[width=1\linewidth]{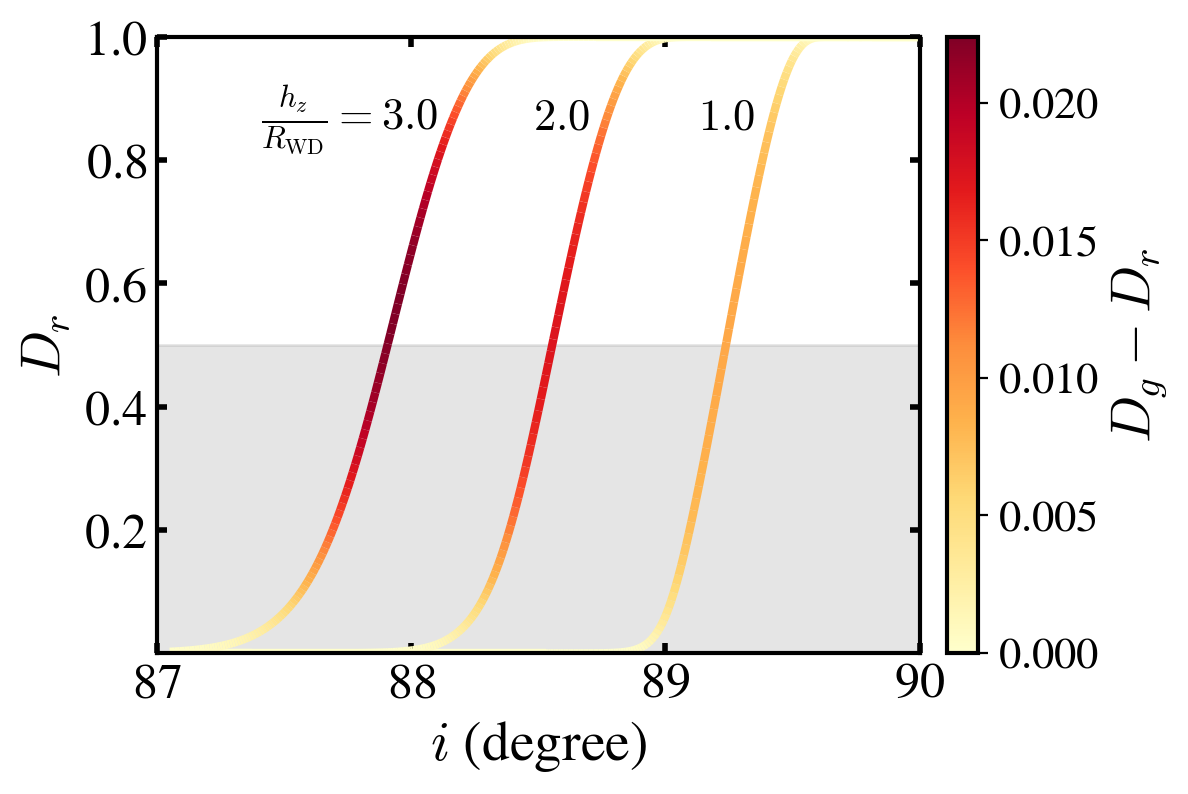}
    \caption{Gray transits from thick disks for WD\,1232$+$563. Same format as Figure \ref{fig:inclination_effect_colorless} but with $\alpha=0.3$, $\tau_{0,~\rm ref}=10^3$, and $a=200~R_{\rm WD}$. I find that my model enables transit over a broad range of inclination angles maintaining the gray nature within the derived upper limit of $D_g-D_r\lesssim0.02$.}
    \label{fig:inclination_effect_colorless_1232}
\end{figure}

I now investigate if the flux from the inner rim can explain the excess. The result is shown in Figure~\ref{fig:inner_edge_flux_wd1232}. I find that a significantly larger value of $h_z\gtrsim2.5$ is required to explain the observed flux solely with this radiation component, with a required inclination angle of $i$$\lesssim$$89$~degrees. Such a disk thickness, though unlikely, is not impossible and studies like \cite{Ballering22} hints at such (or even larger) disk heights. The corresponding transit prospects is presented in Figure~\ref{fig:inclination_effect_colorless_1232}. I find that to remain within the gray limit ($D_g-D_r\lesssim0.02$, \citealt{Hermes25}) with the range of $h_z$ and $i$ given by the infrared flux, one requires a low value of $\alpha\lesssim0.3$.

The ``outer rim" from the backwarmed disk can also contribute a significant amount of flux. With $h_z=2.5$ and with a conservative disk temperature of $600$~K yields an edge-off flux of $\approx$$0.01$~mJy. More efficient backwarming may result in more flux, thus reducing the burden on the inner rim to account for the totality of the infrared flux. Overall, even with WD\,1232$+$563 I show that, within a reasonable range of parameters, the thick thick radiation components can account for the infrared excess and the transits consistently. 

\section{Effect of Inner-Rim Flux on Disk Radius}\label{app:param_study}

The analysis in Section \ref{subsec:inner_rim_ir_flux} and Appendix \ref{app:wd1232} already shows how the flux from the inner rim changes as a function of $h_z$. The spectral shape (in the \citealt{Dullemond03} formulation) is a simple blackbody and, thus, the variation with temperature is well known. The last important parameter is $R_{\rm rim}$. We adopt the parameters for WD\,1145$+$017 for comparison. To show the effect of this parameter on the flux, I consider two cases. First, I keep $T_{\rm rim}$ fixed (at the value for WD\,1145$+$017 of $1145$~K) and vary $R_{\rm rim}$. This case is analogous to fixing the inner rim temperature at a sublimation temperature and varying the white dwarf temperature. In the second case, I vary $R_{\rm rim}$ but consistently calculate the approximate equilibrium temperature (using Equation \ref{eq:trim} with $n=1$ and dropping the cross section correction) adopting the temperature of WD\,1145$+$017 of $T_{\rm WD}=15020$~K. In both cases, I consider $h_z/R_{\rm WD}=0.75$. I only consider the maximum flux, as the variation with inclination angle is also discussed in the previous sections.

\begin{figure}[t]
    \centering
    \includegraphics[width=1\linewidth]{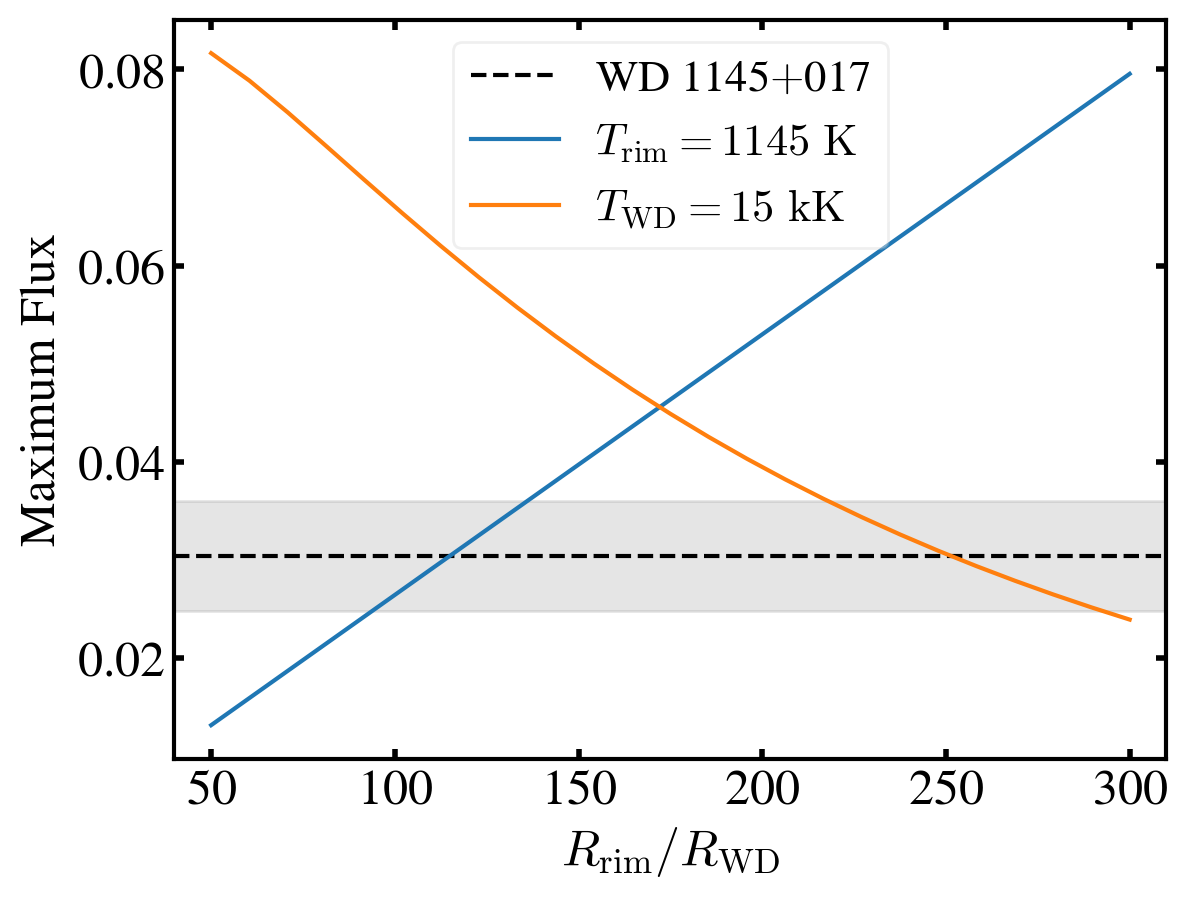}
    \caption{The variation of the maximum flux from the inner rim as a function of $R_{\rm rim}$ in two cases: fixing $T_{\rm rim}$ (blue line) and fixing $T_{\rm WD}$ (orange line) at the WISE W1 wavelength. For comparison, I also shade the W1 flux for WD\,1145$+$017. }
    \label{fig:inner_edge_flux_param_study}
\end{figure}

The result in shown in Figure~\ref{fig:inner_edge_flux_param_study} for the WISE W1 band. I see that when $T_{\rm rim}$ is fixed, the flux grows almost linearly with $R_{\rm rim}$, as the area of the cylinder increases with distance. However, when $T_{\rm rim}$ is allowed to vary, the flux decreases with increasing disk radius. This is because the effect in the drop of temperature exceeds that of the increment in area. Note that this is a simplistic treatment of this complex radiation component. Detailed modeling is required to understand the parameter dependence more accurately.

\section{Effect of Disk Parameters on Optically-thin Emission}\label{app:si_em_par_study}

\begin{figure*}
    \centering
    \includegraphics[width=1\linewidth]{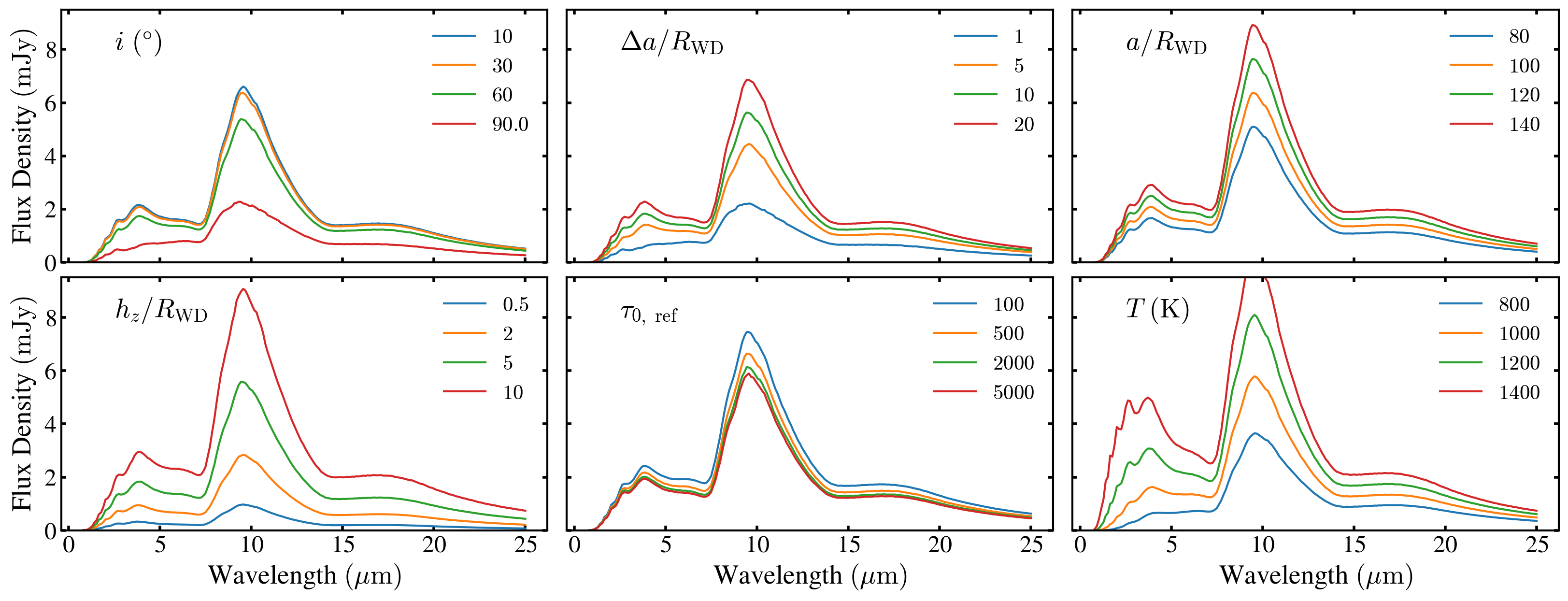}
    \caption{The variation of the optically thin emission flux as the various disk parameters are varied. The base parameter values are the same as opted for G\,29-38 in Section~\ref{subsec:g29-38}: $i=30$ degrees, $\Delta a=15~R_{\rm WD}$, $a=100~R_{\rm WD}$, $h_z=6~R_{\rm WD}$, $\tau_{0,~\rm ref}=10^3$, and $T_s=1090~\rm K$. In each panel, I vary one of the parameters (labeled) by keeping the others fixed at these base values.}
    \label{fig:si_em_feature_par}
\end{figure*}

Here I present how the disk parameters affect the optically thin emission in my model. I use the same parameter values as discussed for G\,29-38 in Section~\ref{subsec:g29-38} as the base values, and then vary each of the parameters individually. Figure~\ref{fig:si_em_feature_par} shows the result of this exercise. Firstly, the emission flux decreases as the inclination is increased. This is because higher $i$ increases the horizontal optical path length (thus optical depth), thus decreasing the amount of optically thin dust. This, thus, serves as a prediction for high-inclination transiting debris systems to have relatively weaker silicate emission feature. Observations with JWST is essential to confirm this, and, along with the near IR excess, will provide essential constraints on the disk geometry. Next, we see that increasing the disk size ($a$, $\Delta a$, or $h_z$) increases the emission flux. This is expected as larger disk dimensions increases the amount of optically thin dust. Increasing $\tau_{0,~\rm ref}$ decreases the emission for obvious reasons. Finally, increasing the dust temperature naturally increases the emission flux proportionately as the blackbody function.


\bibliography{ref}{}

\begin{thebibliography}{}
\expandafter\ifx\csname natexlab\endcsname\relax\def\natexlab#1{#1}\fi
\providecommand{\url}[1]{\href{#1}{#1}}
\providecommand{\dodoi}[1]{doi:~\href{http://doi.org/#1}{\nolinkurl{#1}}}
\providecommand{\doeprint}[1]{\href{http://ascl.net/#1}{\nolinkurl{http://ascl.net/#1}}}
\providecommand{\doarXiv}[1]{\href{https://arxiv.org/abs/#1}{\nolinkurl{https://arxiv.org/abs/#1}}}

\bibitem[{{Alonso} {et~al.}(2016){Alonso}, {Rappaport}, {Deeg}, \& {Palle}}]{Alonso16}
{Alonso}, R., {Rappaport}, S., {Deeg}, H.~J., \& {Palle}, E. 2016, \aap, 589, L6, \dodoi{10.1051/0004-6361/201628511}

\bibitem[{{Astropy Collaboration} {et~al.}(2013){Astropy Collaboration}, {Robitaille}, {Tollerud}, {Greenfield}, {Droettboom}, {Bray}, {Aldcroft}, {Davis}, {Ginsburg}, {Price-Whelan}, {Kerzendorf}, {Conley}, {Crighton}, {Barbary}, {Muna}, {Ferguson}, {Grollier}, {Parikh}, {Nair}, {Unther}, {Deil}, {Woillez}, {Conseil}, {Kramer}, {Turner}, {Singer}, {Fox}, {Weaver}, {Zabalza}, {Edwards}, {Azalee Bostroem}, {Burke}, {Casey}, {Crawford}, {Dencheva}, {Ely}, {Jenness}, {Labrie}, {Lim}, {Pierfederici}, {Pontzen}, {Ptak}, {Refsdal}, {Servillat}, \& {Streicher}}]{Astropy13}
{Astropy Collaboration}, {Robitaille}, T.~P., {Tollerud}, E.~J., {et~al.} 2013, \aap, 558, A33, \dodoi{10.1051/0004-6361/201322068}

\bibitem[{{Astropy Collaboration} {et~al.}(2018){Astropy Collaboration}, {Price-Whelan}, {Sip{\H{o}}cz}, {G{\"u}nther}, {Lim}, {Crawford}, {Conseil}, {Shupe}, {Craig}, {Dencheva}, {Ginsburg}, {VanderPlas}, {Bradley}, {P{\'e}rez-Su{\'a}rez}, {de Val-Borro}, {Aldcroft}, {Cruz}, {Robitaille}, {Tollerud}, {Ardelean}, {Babej}, {Bach}, {Bachetti}, {Bakanov}, {Bamford}, {Barentsen}, {Barmby}, {Baumbach}, {Berry}, {Biscani}, {Boquien}, {Bostroem}, {Bouma}, {Brammer}, {Bray}, {Breytenbach}, {Buddelmeijer}, {Burke}, {Calderone}, {Cano Rodr{\'\i}guez}, {Cara}, {Cardoso}, {Cheedella}, {Copin}, {Corrales}, {Crichton}, {D'Avella}, {Deil}, {Depagne}, {Dietrich}, {Donath}, {Droettboom}, {Earl}, {Erben}, {Fabbro}, {Ferreira}, {Finethy}, {Fox}, {Garrison}, {Gibbons}, {Goldstein}, {Gommers}, {Greco}, {Greenfield}, {Groener}, {Grollier}, {Hagen}, {Hirst}, {Homeier}, {Horton}, {Hosseinzadeh}, {Hu}, {Hunkeler}, {Ivezi{\'c}}, {Jain}, {Jenness}, {Kanarek}, {Kendrew}, {Kern}, {Kerzendorf}, {Khvalko}, {King}, {Kirkby}, {Kulkarni},
  {Kumar}, {Lee}, {Lenz}, {Littlefair}, {Ma}, {Macleod}, {Mastropietro}, {McCully}, {Montagnac}, {Morris}, {Mueller}, {Mumford}, {Muna}, {Murphy}, {Nelson}, {Nguyen}, {Ninan}, {N{\"o}the}, {Ogaz}, {Oh}, {Parejko}, {Parley}, {Pascual}, {Patil}, {Patil}, {Plunkett}, {Prochaska}, {Rastogi}, {Reddy Janga}, {Sabater}, {Sakurikar}, {Seifert}, {Sherbert}, {Sherwood-Taylor}, {Shih}, {Sick}, {Silbiger}, {Singanamalla}, {Singer}, {Sladen}, {Sooley}, {Sornarajah}, {Streicher}, {Teuben}, {Thomas}, {Tremblay}, {Turner}, {Terr{\'o}n}, {van Kerkwijk}, {de la Vega}, {Watkins}, {Weaver}, {Whitmore}, {Woillez}, {Zabalza}, \& {Astropy Contributors}}]{Astropy18}
{Astropy Collaboration}, {Price-Whelan}, A.~M., {Sip{\H{o}}cz}, B.~M., {et~al.} 2018, \aj, 156, 123, \dodoi{10.3847/1538-3881/aabc4f}

\bibitem[{{Ballering} {et~al.}(2022){Ballering}, {Levens}, {Su}, \& {Cleeves}}]{Ballering22}
{Ballering}, N.~P., {Levens}, C.~I., {Su}, K. Y.~L., \& {Cleeves}, L.~I. 2022, \apj, 939, 108, \dodoi{10.3847/1538-4357/ac9a4a}

\bibitem[{{Bhattacharjee} {et~al.}(2025){Bhattacharjee}, {Vanderbosch}, {Hollands}, {Tremblay}, {Xu}, {Guidry}, {Hermes}, {Caiazzo}, {Rodriguez}, {van Roestel}, {Roulston}, {Riddle}, {Rusholme}, {Groom}, {Smith}, \& {Toloza}}]{Bhattacharjee25}
{Bhattacharjee}, S., {Vanderbosch}, Z.~P., {Hollands}, M.~A., {et~al.} 2025, arXiv e-prints, arXiv:2502.05502, \dodoi{10.48550/arXiv.2502.05502}

\bibitem[{{Bonsor} {et~al.}(2017){Bonsor}, {Farihi}, {Wyatt}, \& {van Lieshout}}]{Bonsor17}
{Bonsor}, A., {Farihi}, J., {Wyatt}, M.~C., \& {van Lieshout}, R. 2017, \mnras, 468, 154, \dodoi{10.1093/mnras/stx425}

\bibitem[{{Brouwers} {et~al.}(2022{\natexlab{a}}){Brouwers}, {Bonsor}, \& {Malamud}}]{Brouers22}
{Brouwers}, M.~G., {Bonsor}, A., \& {Malamud}, U. 2022{\natexlab{a}}, \mnras, 509, 2404, \dodoi{10.1093/mnras/stab3009}

\bibitem[{{Brouwers} {et~al.}(2022{\natexlab{b}}){Brouwers}, {Bonsor}, \& {Malamud}}]{Brouwers22}
---. 2022{\natexlab{b}}, \mnras, 509, 2404, \dodoi{10.1093/mnras/stab3009}

\bibitem[{{Calvet} {et~al.}(1991){Calvet}, {Patino}, {Magris}, \& {D'Alessio}}]{Calvet91}
{Calvet}, N., {Patino}, A., {Magris}, G.~C., \& {D'Alessio}, P. 1991, \apj, 380, 617, \dodoi{10.1086/170618}

\bibitem[{{Chiang} \& {Goldreich}(1997)}]{Chiang97}
{Chiang}, E.~I., \& {Goldreich}, P. 1997, \apj, 490, 368, \dodoi{10.1086/304869}

\bibitem[{{Chrenko} {et~al.}(2024){Chrenko}, {Flock}, {Ueda}, {M{\'e}rand}, {Benisty}, \& {Chametla}}]{Chrenko24}
{Chrenko}, O., {Flock}, M., {Ueda}, T., {et~al.} 2024, \aj, 167, 124, \dodoi{10.3847/1538-3881/ad234d}

\bibitem[{{Croll} {et~al.}(2014){Croll}, {Rappaport}, {DeVore}, {Gilliland}, {Crepp}, {Howard}, {Star}, {Chiang}, {Levine}, {Jenkins}, {Albert}, {Bonomo}, {Fortney}, \& {Isaacson}}]{Croll14}
{Croll}, B., {Rappaport}, S., {DeVore}, J., {et~al.} 2014, \apj, 786, 100, \dodoi{10.1088/0004-637X/786/2/100}

\bibitem[{{Debes} {et~al.}(2011){Debes}, {Hoard}, {Wachter}, {Leisawitz}, \& {Cohen}}]{Debes11}
{Debes}, J.~H., {Hoard}, D.~W., {Wachter}, S., {Leisawitz}, D.~T., \& {Cohen}, M. 2011, \apjs, 197, 38, \dodoi{10.1088/0067-0049/197/2/38}

\bibitem[{{Debes} \& {Sigurdsson}(2002)}]{Debes02}
{Debes}, J.~H., \& {Sigurdsson}, S. 2002, \apj, 572, 556, \dodoi{10.1086/340291}

\bibitem[{{Draine}(2003)}]{Draine03b}
{Draine}, B.~T. 2003, \apj, 598, 1026, \dodoi{10.1086/379123}

\bibitem[{{Draine}(2011)}]{Draine11}
---. 2011, {Physics of the Interstellar and Intergalactic Medium}

\bibitem[{{Dullemond} {et~al.}(2001){Dullemond}, {Dominik}, \& {Natta}}]{Dullemond03}
{Dullemond}, C.~P., {Dominik}, C., \& {Natta}, A. 2001, \apj, 560, 957, \dodoi{10.1086/323057}

\bibitem[{{Dullemond} \& {Monnier}(2010)}]{Dullemond10}
{Dullemond}, C.~P., \& {Monnier}, J.~D. 2010, \araa, 48, 205, \dodoi{10.1146/annurev-astro-081309-130932}

\bibitem[{{Farihi}(2016)}]{Farihi16}
{Farihi}, J. 2016, \nar, 71, 9, \dodoi{10.1016/j.newar.2016.03.001}

\bibitem[{{Farihi} {et~al.}(2025){Farihi}, {Su}, {Melis}, {Kenyon}, {Swan}, {Redfield}, {Wyatt}, \& {Debes}}]{Farihi25}
{Farihi}, J., {Su}, K.~Y.~L., {Melis}, C., {et~al.} 2025, \apjl, 981, L5, \dodoi{10.3847/2041-8213/adae88}

\bibitem[{{Farihi} {et~al.}(2008){Farihi}, {Zuckerman}, \& {Becklin}}]{Farihi08}
{Farihi}, J., {Zuckerman}, B., \& {Becklin}, E.~E. 2008, \apj, 674, 431, \dodoi{10.1086/521715}

\bibitem[{{Farihi} {et~al.}(2018){Farihi}, {van Lieshout}, {Cauley}, {Dennihy}, {Su}, {Kenyon}, {Wilson}, {Toloza}, {G{\"a}nsicke}, {von Hippel}, {Redfield}, {Debes}, {Xu}, {Rogers}, {Bonsor}, {Swan}, {Pala}, \& {Reach}}]{Farihi18}
{Farihi}, J., {van Lieshout}, R., {Cauley}, P.~W., {et~al.} 2018, \mnras, 481, 2601, \dodoi{10.1093/mnras/sty2331}

\bibitem[{{Farihi} {et~al.}(2022){Farihi}, {Hermes}, {Marsh}, {Mustill}, {Wyatt}, {Guidry}, {Wilson}, {Redfield}, {Izquierdo}, {Toloza}, {G{\"a}nsicke}, {Aungwerojwit}, {Kaewmanee}, {Dhillon}, \& {Swan}}]{Farihi22}
{Farihi}, J., {Hermes}, J.~J., {Marsh}, T.~R., {et~al.} 2022, \mnras, 511, 1647, \dodoi{10.1093/mnras/stab3475}

\bibitem[{{Flock} {et~al.}(2016){Flock}, {Fromang}, {Turner}, \& {Benisty}}]{Flock16}
{Flock}, M., {Fromang}, S., {Turner}, N.~J., \& {Benisty}, M. 2016, \apj, 827, 144, \dodoi{10.3847/0004-637X/827/2/144}

\bibitem[{{Ginsburg} {et~al.}(2019){Ginsburg}, {Sip{\H{o}}cz}, {Brasseur}, {Cowperthwaite}, {Craig}, {Deil}, {Guillochon}, {Guzman}, {Liedtke}, {Lian Lim}, {Lockhart}, {Mommert}, {Morris}, {Norman}, {Parikh}, {Persson}, {Robitaille}, {Segovia}, {Singer}, {Tollerud}, {de Val-Borro}, {Valtchanov}, {Woillez}, {Astroquery Collaboration}, \& {a subset of astropy Collaboration}}]{astroquery19}
{Ginsburg}, A., {Sip{\H{o}}cz}, B.~M., {Brasseur}, C.~E., {et~al.} 2019, \aj, 157, 98, \dodoi{10.3847/1538-3881/aafc33}

\bibitem[{{Guidry} {et~al.}(2024){Guidry}, {Hermes}, {De}, {Ould Rouis}, {Ewing}, \& {Kaiser}}]{Guidry24b}
{Guidry}, J.~A., {Hermes}, J.~J., {De}, K., {et~al.} 2024, \apj, 972, 126, \dodoi{10.3847/1538-4357/ad5be7}

\bibitem[{{Guidry} {et~al.}(2021){Guidry}, {Vanderbosch}, {Hermes}, {Barlow}, {Lopez}, {Boudreaux}, {Corcoran}, {Bell}, {Montgomery}, {Heintz}, {Castanheira}, {Reding}, {Dunlap}, {Winget}, {Winget}, \& {Kuehne}}]{Guidry21}
{Guidry}, J.~A., {Vanderbosch}, Z.~P., {Hermes}, J.~J., {et~al.} 2021, \apj, 912, 125, \dodoi{10.3847/1538-4357/abee68}

\bibitem[{{Hallakoun} {et~al.}(2017){Hallakoun}, {Xu}, {Maoz}, {Marsh}, {Ivanov}, {Dhillon}, {Bours}, {Parsons}, {Kerry}, {Sharma}, {Su}, {Rengaswamy}, {Pravec}, {Ku{\v{s}}nir{\'a}k}, {Ku{\v{c}}{\'a}kov{\'a}}, {Armstrong}, {Arnold}, {Gerard}, \& {Vanzi}}]{Hallakoun17}
{Hallakoun}, N., {Xu}, S., {Maoz}, D., {et~al.} 2017, \mnras, 469, 3213, \dodoi{10.1093/mnras/stx924}

\bibitem[{{Hansen}(1971)}]{Hansen71}
{Hansen}, J.~E. 1971, Journal of the Atmospheric Sciences, 28, 1400, \dodoi{10.1175/1520-0469(1971)028<1400:MSOPLI>2.0.CO;2}

\bibitem[{Harris {et~al.}(2020)Harris, Millman, van~der Walt, Gommers, Virtanen, Cournapeau, Wieser, Taylor, Berg, Smith, Kern, Picus, Hoyer, van Kerkwijk, Brett, Haldane, del R{\'{i}}o, Wiebe, Peterson, G{\'{e}}rard-Marchant, Sheppard, Reddy, Weckesser, Abbasi, Gohlke, \& Oliphant}]{harris2020array}
Harris, C.~R., Millman, K.~J., van~der Walt, S.~J., {et~al.} 2020, Nature, 585, 357, \dodoi{10.1038/s41586-020-2649-2}

\bibitem[{{Hermes} {et~al.}(2025){Hermes}, {Guidry}, {Vanderbosch}, {Badenas-Agusti}, {Xu}, {Kao}, {Rodriguez}, \& {Hawkins}}]{Hermes25}
{Hermes}, J.~J., {Guidry}, J.~A., {Vanderbosch}, Z.~P., {et~al.} 2025, arXiv e-prints, arXiv:2501.02050, \dodoi{10.48550/arXiv.2501.02050}

\bibitem[{Hunter(2007)}]{Hunter:2007}
Hunter, J.~D. 2007, Computing in Science \& Engineering, 9, 90, \dodoi{10.1109/MCSE.2007.55}

\bibitem[{{Isella} \& {Natta}(2005)}]{Isella05}
{Isella}, A., \& {Natta}, A. 2005, \aap, 438, 899, \dodoi{10.1051/0004-6361:20052773}

\bibitem[{{Izquierdo} {et~al.}(2018){Izquierdo}, {Rodr{\'\i}guez-Gil}, {G{\"a}nsicke}, {Mustill}, {Toloza}, {Tremblay}, {Wyatt}, {Chote}, {Eggl}, {Farihi}, {Koester}, {Lyra}, {Manser}, {Marsh}, {Pall{\'e}}, {Raddi}, {Veras}, {Villaver}, \& {Portegies Zwart}}]{Izuierdo18}
{Izquierdo}, P., {Rodr{\'\i}guez-Gil}, P., {G{\"a}nsicke}, B.~T., {et~al.} 2018, \mnras, 481, 703, \dodoi{10.1093/mnras/sty2315}

\bibitem[{{Jura}(2003)}]{Jura03}
{Jura}, M. 2003, \apjl, 584, L91, \dodoi{10.1086/374036}

\bibitem[{{Jura} {et~al.}(2007){Jura}, {Farihi}, \& {Zuckerman}}]{Jura07}
{Jura}, M., {Farihi}, J., \& {Zuckerman}, B. 2007, \apj, 663, 1285, \dodoi{10.1086/518767}

\bibitem[{{Jura} {et~al.}(2009){Jura}, {Farihi}, \& {Zuckerman}}]{Jura09}
---. 2009, \aj, 137, 3191, \dodoi{10.1088/0004-6256/137/2/3191}

\bibitem[{{Kama} {et~al.}(2009){Kama}, {Min}, \& {Dominik}}]{Kama09}
{Kama}, M., {Min}, M., \& {Dominik}, C. 2009, \aap, 506, 1199, \dodoi{10.1051/0004-6361/200912068}

\bibitem[{{Kenyon} \& {Bromley}(2017)}]{Kenyon17}
{Kenyon}, S.~J., \& {Bromley}, B.~C. 2017, \apj, 844, 116, \dodoi{10.3847/1538-4357/aa7b85}

\bibitem[{{Koester} {et~al.}(2014){Koester}, {G{\"a}nsicke}, \& {Farihi}}]{Koester14}
{Koester}, D., {G{\"a}nsicke}, B.~T., \& {Farihi}, J. 2014, \aap, 566, A34, \dodoi{10.1051/0004-6361/201423691}

\bibitem[{{Li} {et~al.}(2025){Li}, {Bonsor}, \& {Shorttle}}]{Li25}
{Li}, Y., {Bonsor}, A., \& {Shorttle}, O. 2025, \mnras, 541, 610, \dodoi{10.1093/mnras/staf1028}

\bibitem[{{Marocco} {et~al.}(2021){Marocco}, {Eisenhardt}, {Fowler}, {Kirkpatrick}, {Meisner}, {Schlafly}, {Stanford}, {Garcia}, {Caselden}, {Cushing}, {Cutri}, {Faherty}, {Gelino}, {Gonzalez}, {Jarrett}, {Koontz}, {Mainzer}, {Marchese}, {Mobasher}, {Schlegel}, {Stern}, {Teplitz}, \& {Wright}}]{Morocco21}
{Marocco}, F., {Eisenhardt}, P. R.~M., {Fowler}, J.~W., {et~al.} 2021, \apjs, 253, 8, \dodoi{10.3847/1538-4365/abd805}

\bibitem[{{Mathis} {et~al.}(1977){Mathis}, {Rumpl}, \& {Nordsieck}}]{Mathis77}
{Mathis}, J.~S., {Rumpl}, W., \& {Nordsieck}, K.~H. 1977, \apj, 217, 425, \dodoi{10.1086/155591}

\bibitem[{{Mullally} {et~al.}(2007){Mullally}, {Kilic}, {Reach}, {Kuchner}, {von Hippel}, {Burrows}, \& {Winget}}]{Mullally07}
{Mullally}, F., {Kilic}, M., {Reach}, W.~T., {et~al.} 2007, \apjs, 171, 206, \dodoi{10.1086/511858}

\bibitem[{{Ould Rouis} {et~al.}(2024){Ould Rouis}, {Hermes}, {G{\"a}nsicke}, {Sahu}, {Koester}, {Tremblay}, {Veras}, {Farihi}, {Heintz}, {Gentile Fusillo}, \& {Redfield}}]{OuldRouis24}
{Ould Rouis}, L.~B., {Hermes}, J.~J., {G{\"a}nsicke}, B.~T., {et~al.} 2024, arXiv e-prints, arXiv:2410.06335, \dodoi{10.48550/arXiv.2410.06335}

\bibitem[{pandas~development team(2020)}]{reback2020pandas}
pandas~development team, T. 2020, pandas-dev/pandas: Pandas, latest,  Zenodo, \dodoi{10.5281/zenodo.3509134}

\bibitem[{Prahl(2023)}]{miepython23}
Prahl, S. 2023, miepython: Pure python implementation of Mie scattering, v2.5.3,  Zenodo, \dodoi{10.5281/zenodo.8218010}

\bibitem[{{Reach} {et~al.}(2009){Reach}, {Lisse}, {von Hippel}, \& {Mullally}}]{Reach09}
{Reach}, W.~T., {Lisse}, C., {von Hippel}, T., \& {Mullally}, F. 2009, \apj, 693, 697, \dodoi{10.1088/0004-637X/693/1/697}

\bibitem[{{Rogers} {et~al.}(2020){Rogers}, {Xu}, {Bonsor}, {Hodgkin}, {Su}, {von Hippel}, \& {Jura}}]{Rogers20}
{Rogers}, L.~K., {Xu}, S., {Bonsor}, A., {et~al.} 2020, \mnras, 494, 2861, \dodoi{10.1093/mnras/staa873}

\bibitem[{{Swan} {et~al.}(2019){Swan}, {Farihi}, \& {Wilson}}]{Swan19}
{Swan}, A., {Farihi}, J., \& {Wilson}, T.~G. 2019, \mnras, 484, L109, \dodoi{10.1093/mnrasl/slz014}

\bibitem[{{Swan} {et~al.}(2020){Swan}, {Farihi}, {Wilson}, \& {Parsons}}]{Swan20}
{Swan}, A., {Farihi}, J., {Wilson}, T.~G., \& {Parsons}, S.~G. 2020, \mnras, 496, 5233, \dodoi{10.1093/mnras/staa1688}

\bibitem[{{Swan} {et~al.}(2021){Swan}, {Kenyon}, {Farihi}, {Dennihy}, {G{\"a}nsicke}, {Hermes}, {Melis}, \& {von Hippel}}]{Swan21}
{Swan}, A., {Kenyon}, S.~J., {Farihi}, J., {et~al.} 2021, \mnras, 506, 432, \dodoi{10.1093/mnras/stab1738}

\bibitem[{{Vanderbosch} {et~al.}(2020){Vanderbosch}, {Hermes}, {Dennihy}, {Dunlap}, {Izquierdo}, {Tremblay}, {Cho}, {G{\"a}nsicke}, {Toloza}, {Bell}, {Montgomery}, \& {Winget}}]{Vanderbosch20}
{Vanderbosch}, Z., {Hermes}, J.~J., {Dennihy}, E., {et~al.} 2020, \apj, 897, 171, \dodoi{10.3847/1538-4357/ab9649}

\bibitem[{{Vanderbosch} {et~al.}(2021){Vanderbosch}, {Rappaport}, {Guidry}, {Gary}, {Blouin}, {Kaye}, {Weinberger}, {Melis}, {Klein}, {Zuckerman}, {Vanderburg}, {Hermes}, {Hegedus}, {Burleigh}, {Sefako}, {Worters}, \& {Heintz}}]{Vanderbosch21}
{Vanderbosch}, Z.~P., {Rappaport}, S., {Guidry}, J.~A., {et~al.} 2021, \apj, 917, 41, \dodoi{10.3847/1538-4357/ac0822}

\bibitem[{{Vanderburg} {et~al.}(2015){Vanderburg}, {Johnson}, {Rappaport}, {Bieryla}, {Irwin}, {Lewis}, {Kipping}, {Brown}, {Dufour}, {Ciardi}, {Angus}, {Schaefer}, {Latham}, {Charbonneau}, {Beichman}, {Eastman}, {McCrady}, {Wittenmyer}, \& {Wright}}]{Vanderburg15}
{Vanderburg}, A., {Johnson}, J.~A., {Rappaport}, S., {et~al.} 2015, \nat, 526, 546, \dodoi{10.1038/nature15527}

\bibitem[{{Veras} {et~al.}(2017){Veras}, {Carter}, {Leinhardt}, \& {G{\"a}nsicke}}]{Veras17}
{Veras}, D., {Carter}, P.~J., {Leinhardt}, Z.~M., \& {G{\"a}nsicke}, B.~T. 2017, \mnras, 465, 1008, \dodoi{10.1093/mnras/stw2748}

\bibitem[{Virtanen {et~al.}(2020)Virtanen, Gommers, Oliphant, Haberland, Reddy, Cournapeau, Burovski, Peterson, Weckesser, Bright, {van der Walt}, Brett, Wilson, Millman, Mayorov, Nelson, Jones, Kern, Larson, Carey, Polat, Feng, Moore, {VanderPlas}, Laxalde, Perktold, Cimrman, Henriksen, Quintero, Harris, Archibald, Ribeiro, Pedregosa, {van Mulbregt}, \& {SciPy 1.0 Contributors}}]{2020SciPy-NMeth}
Virtanen, P., Gommers, R., Oliphant, T.~E., {et~al.} 2020, Nature Methods, 17, 261, \dodoi{10.1038/s41592-019-0686-2}

\bibitem[{{Wang} {et~al.}(2023){Wang}, {Zhang}, {Wang}, {Zhang}, {Fang}, {Gu}, {Guo}, \& {Jiang}}]{Wang23}
{Wang}, L., {Zhang}, X., {Wang}, J., {et~al.} 2023, \apj, 944, 23, \dodoi{10.3847/1538-4357/acaf5a}

\bibitem[{{Xu} \& {Jura}(2014)}]{Xu14}
{Xu}, S., \& {Jura}, M. 2014, \apjl, 792, L39, \dodoi{10.1088/2041-8205/792/2/L39}

\bibitem[{{Xu} {et~al.}(2020){Xu}, {Lai}, \& {Dennihy}}]{Xu20}
{Xu}, S., {Lai}, S., \& {Dennihy}, E. 2020, \apj, 902, 127, \dodoi{10.3847/1538-4357/abb3fc}

\bibitem[{{Xu} {et~al.}(2018{\natexlab{a}}){Xu}, {Su}, {Rogers}, {Bonsor}, {Olofsson}, {Veras}, {van Lieshout}, {Dufour}, {Green}, {Schlawin}, {Farihi}, {Wilson}, {Wilson}, \& {G{\"a}nsicke}}]{Xu18irvar}
{Xu}, S., {Su}, K. Y.~L., {Rogers}, L.~K., {et~al.} 2018{\natexlab{a}}, \apj, 866, 108, \dodoi{10.3847/1538-4357/aadcfe}

\bibitem[{{Xu} {et~al.}(2018{\natexlab{b}}){Xu}, {Rappaport}, {van Lieshout}, {Vanderburg}, {Gary}, {Hallakoun}, {Ivanov}, {Wyatt}, {DeVore}, {Bayliss}, {Bento}, {Bieryla}, {Cameron}, {Cann}, {Croll}, {Collins}, {Dalba}, {Debes}, {Doyle}, {Dufour}, {Ely}, {Espinoza}, {Joner}, {Jura}, {Kaye}, {McClain}, {Muirhead}, {Palle}, {Panka}, {Provencal}, {Randall}, {Rodriguez}, {Scarborough}, {Sefako}, {Shporer}, {Strickland}, {Zhou}, \& {Zuckerman}}]{Xu18}
{Xu}, S., {Rappaport}, S., {van Lieshout}, R., {et~al.} 2018{\natexlab{b}}, \mnras, 474, 4795, \dodoi{10.1093/mnras/stx3023}

\bibitem[{{Zuckerman} \& {Becklin}(1987)}]{Zuckerman87}
{Zuckerman}, B., \& {Becklin}, E.~E. 1987, \nat, 330, 138, \dodoi{10.1038/330138a0}

\bibitem[{{Zuckerman} {et~al.}(2003){Zuckerman}, {Koester}, {Reid}, \& {H{\"u}nsch}}]{Zuckerman03}
{Zuckerman}, B., {Koester}, D., {Reid}, I.~N., \& {H{\"u}nsch}, M. 2003, \apj, 596, 477, \dodoi{10.1086/377492}

\bibitem[{{Zuckerman} {et~al.}(2010){Zuckerman}, {Melis}, {Klein}, {Koester}, \& {Jura}}]{Zuckerman10}
{Zuckerman}, B., {Melis}, C., {Klein}, B., {Koester}, D., \& {Jura}, M. 2010, \apj, 722, 725, \dodoi{10.1088/0004-637X/722/1/725}

\end{thebibliography}
\bibliographystyle{aasjournal}



\end{document}